\documentclass[english]{article}
\usepackage[T1]{fontenc}
\usepackage[latin9]{inputenc}
\usepackage{geometry}
\usepackage{amsmath}
\usepackage{amssymb}

\makeatletter

\providecommand{\tabularnewline}{\\}

\numberwithin{equation}{section}

\usepackage{jheppub}

\author[a,b]{Tamas Gombor,} 
\author[c]{Charlotte Kristjansen,} 
\author[d]{Vasileios Moustakis,} 
\author[c]{and Xin Qian}
\affiliation[a]{MTA-ELTE ''Momentum'' Integrable Quantum Dynamics Research Group, Eötvös Loránd University,\\ Pázmány Péter sétány 1/A, 1117 Budapest, Hungary}
\affiliation[b]{HUN-REN Wigner Research Centre for Physics, \\Konkoly-Thege Miklós u. 29-33, 1121 Budapest, Hungary}
\affiliation[c]{Niels Bohr International Academy, Niels Bohr Institute, Copenhagen University, \\
Blegdamsvej 17, DK-2100 Copenhagen \O, Denmark}
\affiliation[d]{Department of Mathematics, University of Surrey, Guildford, GU2 7XH, UK}
\emailAdd{gombort@caesar.elte.hu,kristjan@nbi.dk,v.moustakis@surrey.ac.uk,xinqian@nbi.ku.dk}

\makeatother

\usepackage{babel}
\abstract{ 
Invoking a quantum dressing procedure as well as the representation theory of twisted Yangians we derive a number of summation formulas for the overlap between integrable matrix product states and Bethe eigenstates which involve only eigenvalues of fused transfer matrices and which are valid in the presence of inhomogeneities as well as twists.  Although the method is general we specialize to the $SO(6)$ spin chain for which integrable matrix product states corresponding to evaluation representations of the twisted Yangian $Y^+(4)$ encode the information about one-point functions of the D3-D5 domain wall version of ${\cal N}=4$ SYM. Considering the untwisted and homogeneous limit of our summation formulas we finally fill the last gap in the analytical understanding of the overlap formula for the $SO(6$) sector of the D3-D5 domain wall system.

}

\begin{document}
\title{On exact overlaps of integrable matrix product states: inhomogeneities, twists and dressing formulas}
\maketitle

\section{Introduction}

Matrix product states of integrable spin chains have attracted interest as a means to represent boundary states
within the framework of the AdS/CFT correspondence where the boundary state  in the string theory language corresponds
to a probe D-brane and in the gauge theory language to a Nahm pole defect~\cite{deLeeuw:2015hxa,deLeeuw:2016ofj,Kristjansen:2021abc,Kristjansen:2023ysz,deLeeuw:2024qki}. Exploiting
that the good conformal operators of the field theory are in one-to-one correspondence with the  Bethe eigenstates of an underlying integrable
spin chain~\cite{Minahan:2002ve,Minahan:2008hf} the one point function of the defect field theory can be expressed as the overlap between a matrix product state and a Bethe eigenstate~\cite{deLeeuw:2015hxa}.  The technique can also be
used to study one-point function on the Coulomb branch of ${\cal N}=4$ SYM~\cite{Ivanovskiy:2024vel} as well as 
certain three-point functions of the heavy-heavy-light type~\cite{Jiang:2019xdz,Yang:2021hrl,Yang:2022dlk}.

In numerous cases it was possible to find a closed expression for the above mentioned overlaps valid for all Bethe states of the integrable spin chain, see e.g.~\cite{Kristjansen:2024dnm}, and this fact was attributed to the matrix product state being a discrete version of  Zamolodchikov's integrable boundary state~\cite{Zamolodchikov:1989fp,Piroli:2017sei}. 
The overlap 
formulas for integrable matrix product states contain a universal factor which involves the Gaudin matrix of the
Bethe eigenstate being considered. More precisely, the quantity which appears can be viewed as the super determinant of
the Gaudin matrix~\cite{Kristjansen:2020mhn}. The appearance of this structure was observed already in overlap formulas involving simple two-site product states such as the N\'{e}el state in the XXZ spin chain~\cite{Pozsgay_2014,Brockmann_2014odd,Brockmann_2014}.  In addition
to the superdeterminant of the Gaudin matrix overlap formulas for integrable matrix product states contain a non-universal
pre-factor which encodes the properties of the boundary state. For particular types of matrix product states the pre-factor has
a completely factorized form where the factors take the form of Baxter $Q$-functions. These are states built from
repeated two-site states, i.e.\ matrix product states 
which generalize the N\'{e}el state. For the XXZ chain and for general $GL(N)$ spin chains they were studied in respectively~\cite{Gombor:2021uxz} and~\cite{Gombor:2021hmj,Gombor:2023bez} where closed overlap formulas were likewise found.
Such states exist also for integrable super spin chains where the particular matrix product states correspond to field theoretical defects in the limit where the Nahm pole boundary conditions become standard Dirichlet boundary conditions~\cite{Kristjansen:2020mhn,Kristjansen:2020vbe,Kristjansen:2021abc}. 

For more generic, still integrable matrix product states such as the states describing the co-dimension one Nahm pole
defect in ${\cal N}=4$ super Yang-Mills theory it has been observed that the pre-factor is in general not factorized but expressible as a sum over ratios of Baxter $Q$-functions, with the sum apparently being related to fused transfer matrices~\cite{Buhl-Mortensen:2015gfd,DeLeeuw:2018cal}. 
 For concreteness, let us present the matrix product state in question 
\begin{equation}
\langle \mbox{MPS$_k$} | = \sum_{i_1,\ldots i_{2J}} \mbox{Tr}(\omega_{i_{2J}}\ldots\omega_{i_{1}}) \langle e_{i_1}\ldots e_{i_{2J}}|.  \label{intro_eqn1}
\end{equation}
This is a state of the integrable $SO(6)$ spin chain in the fundamental representation, hence  
$i_1,\ldots, i_{2J}\in \{1,2,\ldots, 6\}$. The six matrices involved, when collected in a list $\bar{\omega}=\{\omega_j\}_{j=1}^6$
 take the form
 \begin{equation}
 \bar{\omega}=\{S_1,S_2,S_3,S_3,S_2,S_1\}, \label{intro_eqn2}
 \end{equation}
 where the $S_i$, $i=1,2,3,$ constitute a $k$-dimensional irreducible representation of $\mathfrak{su}(2)$. The basis
 vectors $e_i$, $i=1,\ldots,6$ represent the three complex scalar fields of ${\cal N}=4$
 SYM and their complex conjugates in the following way
 \begin{equation}
 \bar{e}=\{Z,Y,X,\bar{X},\bar{Y},\bar{Z}\}. \label{intro_eqn3}
 \end{equation}
For a detailed description of the set-up in the AdS/CFT language we refer to~\cite{deLeeuw:2017cop,deLeeuw:2019usb,Linardopoulos:2020jck}. 

There exists an easily applicable test to determine whether one can expect a closed overlap formula to exist for a given  matrix product state, consisting in checking whether it is annihilated by all the parity odd conserved charges of its integrable spin chain 
 host~\cite{Piroli:2017sei}. In practice, it is even sufficient to check if this criterion is fulfilled for the first in the 
 series of odd charges~\cite{Buhl-Mortensen:2015gfd,deLeeuw:2016ofj}. In order to actually derive the overlap formula a deeper understanding of the concept of boundary integrability is needed. In the same way as one needs the so-called 
 RTT relation (to be made explicit in the next section) and not just the commutation relation for transfer matrices when one wants to explicitly determine the eigenstates and eigenvectors of an integrable
 spin chain, a relation which we will denote as the KT relation is needed to explicitly derive overlap formulas for the chain with 
a boundary.  It turns out that the appropriate mathematical tool
to deal with this type of relation is that of representation theory of twisted Yangians~\cite{Pozsgay:2018dzs,DeLeeuw:2019ohp}.
The information about the boundary is encoded in a $K$-matrix which has to fulfil the former KT relation, and a $K$-matrix
with this property precisely constitutes a representation of a twisted Yangian~\cite{Molev:1997wp}. The relevance of twisted Yangians for integrable systems with boundaries was understood also in e.g.~\cite{Doikou:2004hy,MacKay:2004tc}. 
Twisted Yangians  have a co-module property which we will exploit to show that given one $K$-matrix which solves the KT relation, one can  perform a dressing by the
Yangian itself which gives rise to a novel solution of the KT relation. This dressing is a quantum analogue of the well-known dressing procedure of~\cite{Zakharov:1973pp,Zakharov:1980ty,Harnad:1983we} by means of which one can generate novel solutions of a classically integrable equation in terms of already known solutions and which has earlier been used in the
AdS/CFT context for generating novel solutions of the classical string equations of motion~\cite{Spradlin:2006wk,Kalousios:2006xy,Jevicki:2007pk}.  
The overlap formula corresponding to the 
one-dimensional representation of the Yangian can be derived by existing methods (at least for vanishing twist) as the corresponding boundary state
constitutes a special example of a two-site product state for which the overlap formula was determined for any GL(N) spin
chain
in~\cite{Gombor:2021hmj,Gombor:2023bez}.
$K$-matrices which
result from dressing the trivial representation give rise to overlaps which can directly be expressed in terms of eigenvalues of
a single transfer matrix. In the present paper we will need to go beyond this type of $K$-matrices.

We have earlier pointed out the importance of the KT relation 
and the twisted Yangian for the computation of spin chain overlaps~\cite{DeLeeuw:2019ohp,Gombor:2021uxz} and we have made use of such considerations to determine the overlap formula for two specific matrix product states of relevance for the study of Nahm pole defects in ${\cal N}=4$ SYM. One of these was a matrix product state involving matrices whose commutators were the
generators of an $\mathfrak{so}$(5) algebra and whose overlap formula encoded the one-point functions in a defect version of ${\cal N}=4$ SYM dual to a D3-D7 probe brane model with flux~\cite{DeLeeuw:2019ohp,Constable:2001ag,Myers:2008me}. The other one was a matrix product
state of the type given in eqn.~(\ref{intro_eqn1}), but restricted to an $\mathfrak{su}(3)$ subsector, meaning that 
one allows only $i_1,\ldots,i_{2J} \in \{1,2,3\}$ and considers the state to 
be a state in an integrable $SU(3)$ spin chain~\cite{deLeeuw:2016umh,DeLeeuw:2019ohp}. 
So far the overlap formula for the complete matrix product state in eqn.~(\ref{intro_eqn2}) has resisted analytical treatment although a beautiful closed expression has been found numerically~\cite{DeLeeuw:2018cal}. 

In this paper we sharpen and extend our  twisted Yangian approach to matrix product states and fill this last gap in the analytical understanding of the D3-D5 overlap formula. In particular, we will allow for inhomogeneities and twists 
which were so far only considered in connection with overlaps for non-nested spin chains~\cite{Gombor:2021uxz}. 
As we will explain the general matrix product state~(\ref{intro_eqn1}) corresponds
to $K$-matrices that constitute so-called evaluation representations of the twisted Yangian $Y^+(4)$. These $K$-matrices
can not be reached directly by dressing the the trivial representation but requires one to go through a recursive 
dressing procedure by means of which one eventually gets an expression for the $K$-matrix as a linear combination of different, dressed versions of the trivial representation. The recursive dressing procedure is slightly different for integer and half-integer
values of the spin and results in two different dressing formulas. These two dressing formulas constitute our main results
and appear in eqns.~(\ref{eq:identity}) and (\ref{eq:halfintdressing}). We stress that as will be explained in the text these formulas are valid also in the presence of twists and inhomogeneities. Considering the limit of vanishing twist and  inhomogeneities we recover the overlap formula obtained numerically in~\cite{DeLeeuw:2018cal} as a sum over eigenvalues  of different transfer 
matrices.\footnote{It should be noted that one can also recover the numerical result by an argument which builds on assuming the full overlap to be a sum over products of two-particle overlaps at the various levels of nesting~\cite{Gombor:2020kgu,Gombor:2020auk,Gombor:2022aqj}. This argument, however, does not reveal  the connection to fused transfer matrices.}

Our paper is organized as follows. We begin in section~\ref{definitions} by  introducing the tools of group theory 
and of integrability needed to study the integrable $\mathfrak{so}(6)$ spin chain with twists and inhomogeneities in its fundamental representation. Subsequently, in section~\ref{Boundary_states} we introduce the $K$-matrix and the crucial KT relation which defines
the concept of an integrable boundary state. Making use of the fact that a $K$-matrix which solves the KT relation constitutes
a representation of the twisted Yangian $Y^+(4)$ we devise a quantum dressing procedure which allows us to generate
a series of integrable matrix product states starting from the trivial representation.
 In section~\ref{Main_Dressing} we turn to deriving, by means of examples, the generalized dressing formulas necessary to reach the two classes of  integrable matrix product states which correspond to evaluation representations of $Y^+(4)$, the representations of relevance for the D3-D5 domain wall set-up. The details of of this derivation are relegated to
appendices~\ref{integerbranching} and~\ref{halfintegerbranching} whereas appendix~\ref{K-matrices}  contains the
explicit $K$-matrices of the evaluation representations. Finally, in section~\ref{D3D5} by considering the limit of vanishing twist and impurities we recover the overlap formula for the general D3-D5 matrix product state given in eqn.~(\ref{intro_eqn1}),
obtained numerically in~\cite{DeLeeuw:2018cal}.  We  also show that our dressing formulas are  in fact valid for any representation
of the quantum space. Section~\ref{conclusion} contains our conclusion.

\section{Definitions for the SO(6) spin chains \label{definitions}}

We begin by giving a few basic definitions related to the algebra $\mathfrak{gl}_{4}$ and the integrability description of 
the $SO(6)$ spin chain.

\subsection{The $\mathfrak{gl}_{4}$ Lie-algebra and its representations}

Let $e_{i}$ be the usual basis in $\mathbb{C}^{4}$ and let $e_{i,j}$
be the unit matrices in $\mathbb{C}^{4}$ defined by $e_{i,j}e_{k}=\delta_{j,k}e_{i}$. Furthermore,
let $e_{[i,j]}$ for $1\leq i<j\leq4$ be a basis of $\mathbb{C}^{6}$
 coming from the anti-symmetrization of $\mathbb{C}^{4}\otimes\mathbb{C}^{4}$ and let
$e_{[i,j],[k,l]}$ be the unit matrices in $\mathbb{C}^{6}$ which fulfil
$e_{[i,j],[k,l]}e_{[a,b]}=\delta_{k,a}\delta_{l,b}e_{[i,j]}$. We
can identify the six dimensional basis with the three complex fields  in ${\cal N}=4$ SYM and their complex conjugates as follows
\begin{align}
Z & \equiv e_{[1,2]}, & \bar{X} & \equiv e_{[2,3]},\nonumber \\
Y & \equiv e_{[1,3]}, & \bar{Y} & \equiv e_{[4,2]},\\
X & \equiv e_{[1,4]}, & \bar{Z} & \equiv e_{[3,4]},\nonumber 
\end{align}
where $e_{[i,j]}=-e_{[j,i]}$.

We also define the $\mathfrak{gl}_{4}$ algebra as
\begin{equation}
\left[E_{i,j},E_{k,l}\right]=\delta_{j,k}E_{i,l}-\delta_{i,l}E_{k,j}.
\end{equation}
The matrices $e_{i,j}$ constitute the defining representation. Let us introduce
the matrices $\mathcal{E}_{i,j}$ 
\begin{equation}
\mathcal{E}_{i,j}=\sum_{c=1}^{4}e_{[i,c],[j,c]},
\end{equation}
which generate the six-dimensional representation. For every 4-tuple
$\lambda\equiv(\lambda_{1},\lambda_{2},\lambda_{3},\lambda_{4})$
we can define a highest weight irreducible representation (irrep)
$E_{i,j}^{\lambda}$ of $\mathfrak{gl}_{4}$ by
\begin{equation}
\begin{split}E_{i,i}^{\lambda}|0_{\lambda}\rangle & =\lambda_{i}|0_{\lambda}\rangle,\\
E_{i,j}^{\lambda}|0_{\lambda}\rangle & =0,\quad\text{for }i<j,
\end{split}
\end{equation}
where $|0_{\lambda}\rangle$ is the highest weight state. We will make
use of the Gelfand-Tsetlin basis of the $\mathfrak{gl}_{4}$ irreps, see e.g.~\cite{molev2002gelfandtsetlinbasesclassicallie}.
Every basis vector $|\Lambda\rangle$ of the irrep $E_{i,j}^{\lambda}$ corresponds to a GT-pattern
\begin{equation}
\Lambda=\begin{array}{ccccccc}
\lambda_{4,1} &  & \lambda_{4,2} &  & \lambda_{4,3} &  & \lambda_{4,4}\\
 & \lambda_{3,1} &  & \lambda_{3,2} &  & \lambda_{3,3}\\
 &  & \lambda_{2,1} &  & \lambda_{2,2}\\
 &  &  & \lambda_{1,1}
\end{array}
\end{equation}
consisting of non-negative integers for which $\lambda_{l+1,k}\geq\lambda_{l,k}\geq\lambda_{l+1,k+1}$. 
The first row contains the highest weights $\lambda_{4,k}\equiv\lambda_{k}$ of the representation.
The weights of the basis vectors $|\Lambda\rangle$  can be expressed as
\begin{equation}
\begin{split}E_{i,i}^{\lambda}|\Lambda\rangle & =\omega_{i}|\Lambda\rangle,\\
\omega_{1} & =\lambda_{1,1},\quad\omega_{j}=\left(\sum_{k=1}^{j}\lambda_{j,k}-\sum_{k=1}^{j-1}\lambda_{j-1,k}\right).
\end{split}
\end{equation}

\subsection{Transfer matrices\label{notation}}

We define the following Lax-operators
\begin{equation}
\begin{split}\mathcal{L}^{\mathbf{4},\mathbf{6}}(u) & =\sum_{i,j=1}^{4}e_{i,j}\otimes\left(\delta_{i,j}+\frac{1}{u+1/2}\mathcal{E}_{j,i}\right),\qquad\widehat{\mathcal{L}}^{\mathbf{4},\mathbf{6}}(u)=\sum_{i,j=1}^{4}e_{5-i,5-j}\otimes\left(\delta_{i,j}+\frac{1}{-u+1/2}\mathcal{E}_{i,j}\right),\\
\mathcal{L}^{\lambda,\mathbf{4}}(u) & =\sum_{i,j=1}^{4}\left(\delta_{i,j}+\frac{1}{u}E_{j,i}^{\lambda}\right)\otimes e_{i,j},\\
\mathcal{L}^{\lambda,\mathbf{6}}(u) & =\sum_{i<j,k<l}\Biggl((u+1/2)(\delta_{i,k}\delta_{j,l}-\delta_{k,j}\delta_{i,l})+\frac{u+1/2}{u-1/2}\left(\delta_{j,l}E_{k,i}^{\lambda}-\delta_{i,l}E_{k,j}^{\lambda}\right)+\left(\delta_{i,k}E_{l,j}^{\lambda}-\delta_{k,j}E_{l,i}^{\lambda}\right)+\\
 & +\frac{1}{u-1/2}\left(E_{k,i}^{\lambda}E_{l,j}^{\lambda}-E_{k,j}^{\lambda}E_{l,i}^{\lambda}\right)\Biggr)\otimes e_{[i,j],[k,l]}.
\end{split}
\end{equation}
The first two act on a product space of the four-dimensional space of the defining representation and the six-dimensional
space carried by the $\mathcal{E}_{i,j}$. For the last two, one of the product spaces involved is the space carried by
the representation $E_{i,j}^{\lambda}$ which is of dimension
\begin{equation}
d_{\lambda}=\frac{1}{12} \,\prod_{i<j}(l_i-l_j), \hspace{0.5cm} l_i=\lambda_i+4-i, \hspace{0.5cm}i=1,\ldots,4.
\end{equation}
The first factor in the product space is referred to as the auxiliary space and the second one as the quantum space.
The Lax matrices above satisfy the following algebra
\begin{equation} \label{RLL}
\begin{split}R_{1,2}(u-v)\mathcal{L}_{1,3}^{\mathbf{4},\mathbf{6}}(u)\mathcal{L}_{2,3}^{\mathbf{4},\mathbf{6}}(v) & =\mathcal{L}_{2,3}^{\mathbf{4},\mathbf{6}}(v)\mathcal{L}_{1,3}^{\mathbf{4},\mathbf{6}}(u)R_{1,2}(u-v),\\
\bar{R}_{1,2}(u-v)\widehat{\mathcal{L}}_{1,3}^{\mathbf{4},\mathbf{6}}(u)\mathcal{L}_{2,3}^{\mathbf{4},\mathbf{6}}(v) & =\mathcal{L}_{2,3}^{\mathbf{4},\mathbf{6}}(v)\widehat{\mathcal{L}}_{1,3}^{\mathbf{4},\mathbf{6}}(u)\bar{R}_{1,2}(u-v),\\
R_{1,2}(u-v)\widehat{\mathcal{L}}_{1,3}^{\mathbf{4},\mathbf{6}}(u)\widehat{\mathcal{L}}_{2,3}^{\mathbf{4},\mathbf{6}}(v) & =\widehat{\mathcal{L}}_{2,3}^{\mathbf{4},\mathbf{6}}(v)\widehat{\mathcal{L}}_{1,3}^{\mathbf{4},\mathbf{6}}(u)R_{1,2}(u-v),\\
\mathcal{L}_{1,2}^{\lambda,\mathbf{4}}(u-v)\mathcal{L}_{1,3}^{\lambda,\mathbf{6}}(u)\mathcal{L}_{2,3}^{\mathbf{4},\mathbf{6}}(v) & =\mathcal{L}_{2,3}^{\mathbf{4},\mathbf{6}}(v)\mathcal{L}_{1,3}^{\lambda,\mathbf{6}}(u)\mathcal{L}_{1,2}^{\lambda,\mathbf{4}}(u-v),
\end{split}
\end{equation}
where we introduced the R-matrices
\begin{equation}
\begin{split}R_{1,2}(u) & =\mathbf{1}+\frac{1}{u}\mathbf{P},\qquad\mathbf{P}=\sum_{i,j=1}^{4}e_{i,j}\otimes e_{j,i},\\
\bar{R}_{1,2}(u) & =\mathbf{1}-\frac{1}{u}\mathbf{Q},\qquad\mathbf{Q}=\sum_{i,j=1}^{4}e_{i,j}\otimes e_{5-i,5-j}.
\end{split}
\end{equation}
These $R$-matrices have the $\mathfrak{gl}_{4}$ symmetry
\begin{equation}
\begin{split}R_{1,2}(u)G_{1}G_{2} & =G_{1}G_{2}R_{1,2}(u),\\
\bar{R}_{1,2}(u)G_{1}(G^{t})^{-1}_{2} & =G_{1}(G^{t})^{-1}_{2}\bar{R}_{1,2}(u),
\end{split}
\end{equation}
where $G\in GL(4)$ and the superscript
$t$ is a special transposition in the auxiliary space, for which $\left[ G^{t}\right]_{i,j}=G_{5-j,5-i}$.

\begin{figure}
\centering
\includegraphics[width=0.95\textwidth]{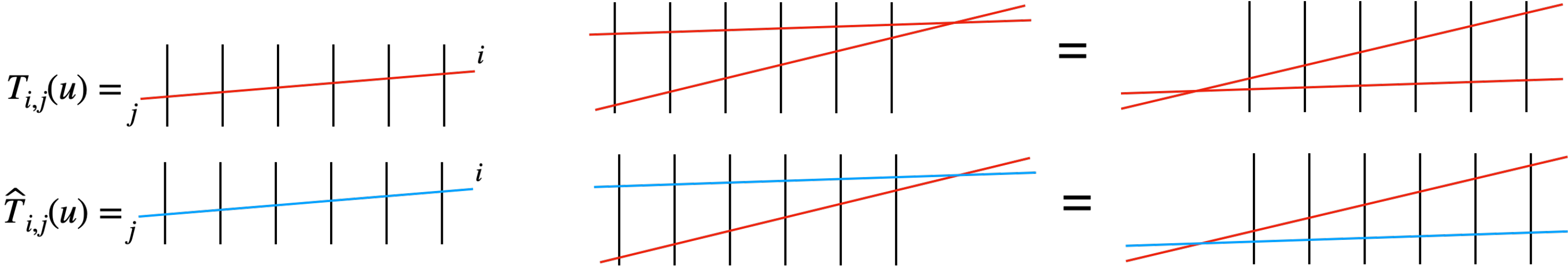} 
\caption{A graphical representation of the monodromy matrices and the RTT-relations. The red and blue lines correspond to the auxiliary space ($i,j=1,\dots 4$). The black lines are the six-dimensional representations. The intersection of two red lines 
represents the R-matrix, $R(u)$, and the  intersection of a red and a blue line the crossed R-matrix, $\bar{R}(u)$.}
\label{fig:RTT}
\end{figure}

Monodromy matrices can be defined in the usual way as follows 
\begin{align}
T_{0}(u) & =\mathcal{L}_{0,2J}^{\mathbf{4},\mathbf{6}}(u+\theta_{J})\mathcal{L}_{0,2J-1}^{\mathbf{4},\mathbf{6}}(u-\theta_{J})\dots\mathcal{L}_{0,2}^{\mathbf{4},\mathbf{6}}(u+\theta_{1})\mathcal{L}_{0,1}^{\mathbf{4},\mathbf{6}}(u-\theta_{1})\nonumber, \\
\widehat{T}_{0}(u) & =\widehat{\mathcal{L}}_{0,2J}^{\mathbf{4},\mathbf{6}}(u+\theta_{J})\widehat{\mathcal{L}}_{0,2J-1}^{\mathbf{4},\mathbf{6}}(u-\theta_{J})\dots\widehat{\mathcal{L}}_{0,2}^{\mathbf{4},\mathbf{6}}(u+\theta_{1})\widehat{\mathcal{L}}_{0,1}^{\mathbf{4},\mathbf{6}}(u-\theta_{1}),\\
T_{0}^{\lambda}(u) & =\mathcal{L}_{0,2J}^{\lambda,\mathbf{6}}(u+\theta_{J})\mathcal{L}_{0,2J-1}^{\lambda,\mathbf{6}}(u-\theta_{J})\dots\mathcal{L}_{0,2}^{\lambda,\mathbf{6}}(u+\theta_{1})\mathcal{L}_{0,1}^{\lambda,\mathbf{6}}(u-\theta_{1}),\nonumber 
\end{align}
where we note that we have introduced inhomogeneities in the form of the $\{\theta_i\}_{i=1}^J$ and that we consider a
system of length $2J$.
The monodromy matrices satisfy the RTT-algebra
\begin{equation}
\begin{split}R_{1,2}(u-v)T_{1}(u)T_{2}(v) & =T_{2}(v)T_{1}(u)R_{1,2}(u-v),\\
\bar{R}_{1,2}(u-v)\widehat{T}_{1}(u)T_{2}(v) & =T_{2}(v)\widehat{T}_{1}(u)\bar{R}_{1,2}(u-v),\\
R_{1,2}(u-v)\widehat{T}_{1}(u)\widehat{T}_{2}(v) & =\widehat{T}_{2}(v)\widehat{T}_{1}(u)R_{1,2}(u-v),\\
\mathcal{L}_{1,2}^{\lambda,\mathbf{4}}(u-v)T_{1}^{\lambda}(u)T_{2}(v) & =T_{2}(v)T_{1}^{\lambda}(u)\mathcal{L}_{1,2}^{\lambda,\mathbf{4}}(u-v).
\end{split}
\label{eq:comm1}
\end{equation}
The $\widehat{T}$ is the ''inverse'' monodromy matrix
\begin{equation}
\widehat{T}^{t}(u)T(u)=\left[\prod_{j=1}^{J}\frac{(u-\theta_{j})^{2}-9/4}{(u-\theta_{j})^{2}-1/4}\frac{(u+\theta_{j})^{2}-9/4}{(u+\theta_{j})^{2}-1/4}\right]\mathbf{1}.
\end{equation} 
The graphical presentations of the monodromy matrices and the RTT-relations are shown in figure \ref{fig:RTT}.

Next, we define the twisted transfer matrices
\begin{equation}
\mathcal{T}(u)=\mathrm{Tr}_{0}T_{0}(u)G_{0},\quad\widehat{\mathcal{T}}(u)=\mathrm{Tr}_{0}\widehat{T}_{0}(u)\widehat{G}_{0},\quad\mathcal{T}_{\lambda}(u)=\mathrm{Tr}_{0}T_{0}^{\lambda}(u)G_{0}^{\lambda},
\end{equation}
where $G\in GL(4)$ and $\widehat{G}=\left(G^{-1}\right)^{t}$,
and we concentrate on diagonal twists, i.e., 
\begin{equation}
G=\mathrm{diag}(z_{1},z_{2},z_{3},z_{4}),\qquad\widehat{G}=\mathrm{diag}(z_{4}^{-1},z_{3}^{-1},z_{2}^{-1},z_{1}^{-1}).
\end{equation}
The $G$ can be also written as
\begin{equation}
G=\exp(\sum_{k=1}^{4}\phi_{k}e_{k,k}),\quad z_{k}=e^{\phi_{k}},
\end{equation}
and similarly, we can express  the twist matrix $G^{\lambda}$ as
\begin{equation}
G^{\lambda}=\exp(\sum_{k=1}^{4}\phi_{k}E_{k,k}^{\lambda}).
\end{equation}
The transfer matrices generate commuting algebras
\begin{equation}
\left[\mathcal{T}_{\lambda}(u),\mathcal{T}_{\lambda'}(u')\right]=0,
\end{equation}
which can be diagonalized simultaneously
\begin{equation}
\mathcal{T}_{\lambda}(u)|\bar{u}\rangle=\tau_{\lambda}(u|\bar{u})|\bar{u}\rangle,
\end{equation}
where $|\bar{u}\rangle$ is the Bethe vector and $\bar{u}\equiv\left\{ \bar{u}^{1},\bar{u}^{2},\bar{u}^{3}\right\} $
denotes the set of Bethe roots $\bar{u}^{j}=\left\{ u_{k}^{j}\right\} _{k=1}^{n_{j}}$.
The Bethe roots should satisfy the Bethe Ansatz equations

\begin{align}
\frac{z_{1}}{z_{2}} & =-\frac{Q_{1}(u_{k}^{1}+1)}{Q_{1}(u_{k}^{1}-1)}\frac{Q_{2}(u_{k}^{1}-1/2)}{Q_{2}(u_{k}^{1}+1/2)},\nonumber \\
\frac{z_{2}}{z_{3}}\frac{Q_{\theta}(u_{k}^{2}+1/2)}{Q_{\theta}(u_{k}^{2}-1/2)} & =-\frac{Q_{2}(u_{k}^{2}+1)}{Q_{2}(u_{k}^{2}-1)}\frac{Q_{1}(u_{k}^{2}-1/2)}{Q_{1}(u_{k}^{2}+1/2)}\frac{Q_{3}(u_{k}^{2}-1/2)}{Q_{3}(u_{k}^{2}+1/2)},\\
\frac{z_{3}}{z_{4}} & =-\frac{Q_{3}(u_{k}^{3}+1)}{Q_{3}(u_{k}^{3}-1)}\frac{Q_{2}(u_{k}^{3}-1/2)}{Q_{2}(u_{k}^{3}+1/2)},\nonumber 
\end{align}
where we introduced the $Q$-functions
\begin{equation}
Q_{\theta}(u)=\prod_{j=1}^{J}(u-\theta_{j})(u+\theta_{j}),\quad Q_{k}(u)=\prod_{j=1}^{n_{k}}(u-u_{j}^{k}),
\end{equation}
including the inhomogeneities $\{\theta_j\}_{j=1}^J$.
The transfer matrix eigenvalues have the form
\begin{align} 
\tau_{\lambda}(u|\bar{u}) & =\sum_{\Lambda}\left[\prod_{j=1}^{4}z_{j}^{\omega_{j}}\right]\frac{Q_{\theta}(u+1/2+\lambda_{2,1})Q_{\theta}(u-1/2+\lambda_{2,2})}{Q_{\theta}(u-1/2)}\prod_{k=1}^{3}\mathcal{F}_{\Lambda}^{(k)}(u|\bar{u}),\label{eq:eig}\\
\mathcal{F}_{\Lambda}^{(k)}(u|\bar{u}) & =\frac{Q_{k}(u+\lambda_{k+1,k+1}-k/2)}{Q_{k}(u+\lambda_{k,k}-k/2)}\prod_{j=1}^{k}\frac{Q_{k}(u+\lambda_{k+1,j}+k/2-j+1)}{Q_{k}(u+\lambda_{k,j}+k/2-j+1)}\prod_{j=1}^{k-1}\frac{Q_{k}(u+\lambda_{k-1,j}+k/2-j)}{Q_{k}(u+\lambda_{k,j}+k/2-j)}.\nonumber 
\end{align}

\section{Integrable Boundary states and Twisted Yangians\label{Boundary_states}}

Our goal is to calculate overlaps,
$\langle \mbox{MPS}|\bar{u}\rangle$,
between integrable matrix product states and on-shell Bethe
eigenstates where the on-shell Bethe states were  defined in the previous
section. The purpose of this section is to define the integrable MPSs.
 We first define integrable
boundary states and corresponding $K$-matrices using the KT-relation
\cite{Gombor:2021hmj,Gombor:2023bez}. The trace of the boundary state
in the boundary space then gives the MPS. $K$-matrices which fulfil the KT relation constitute representations 
of the twisted Yangian $Y^+(4)$. We briefly review the properties of the evaluation representations of 
$Y^+(4)$~\cite{Molev:1997wp} 
 and introduce the corresponding
MPSs.  These are the types of MPSs which appear in the D3-D5 domain wall problem.

\subsection{Definition of the KT-relation}

\begin{figure}
\centering
\includegraphics[width=0.65\textwidth]{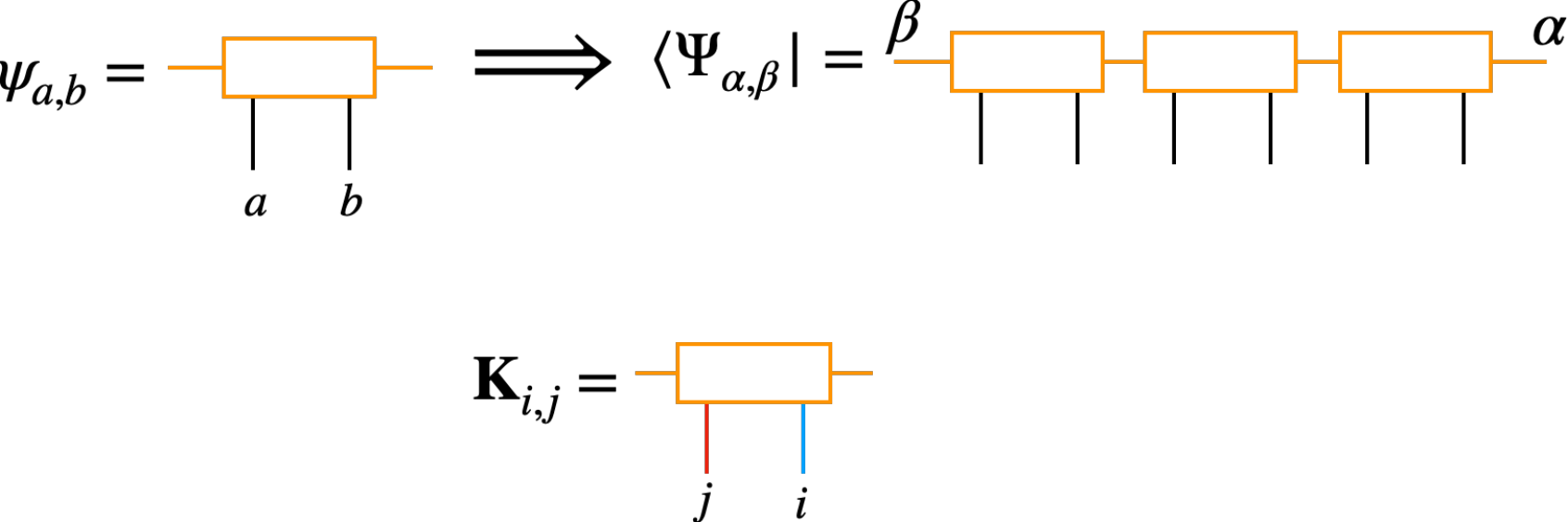} 
\caption{The graphical presentation of the $K$-matrix and the boundary state. The red and blue lines correspond to the auxiliary space ($i,j=1,\dots 4$). The black lines are the six-dimensional representations ($a,b=1,\dots 6$). The boundary space is denoted by orange lines ($\alpha,\beta =1,\dots, d_B$).}
\label{fig:KPsi}
\end{figure}

\begin{figure}
\centering
\includegraphics[width=0.9\textwidth]{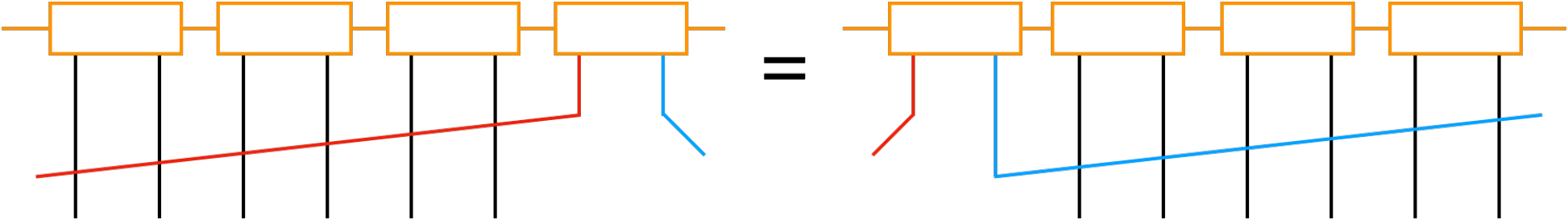} 
\caption{The graphical presentation of the KT-relation. The red and blue lines correspond to the auxiliary space. The black lines are the six-dimensional representations. The boundary space is denoted by orange lines.}
\label{fig:KT}
\end{figure}

The integrable boundary states (which are considered in this paper)
are defined by the crossed KT-relation
\begin{equation}\label{KT}
\sum_{k,\gamma}K_{i,k}^{\alpha,\gamma}(u)\langle\Psi_{\gamma,\beta}|T_{k,j}(u)=\sum_{k,\gamma}\langle\Psi_{\alpha,\gamma}|\widehat{T}_{i,k}(-u)K_{k,j}^{\gamma,\beta}(u),
\end{equation}
where $i,j,k=1,\dots,4$ and refer to the auxiliary space. The indices  $\alpha,\beta,\gamma$ take values in the set $\{1,\ldots ,d_{B}\}$,
and refer to a boundary vector space $\mathcal{V}_{B}\cong\mathbb{C}^{d_{B}}$ of dimension $d_{B}$, 
denoted as the bond dimension.
The states $\langle\Psi_{\alpha,\beta}|$ 
are
 covectors in the quantum space, $\mathcal{H}$,  and  $K_{i,k}^{\alpha,\gamma}(u)$'s are scalars in this space.
We can thus define matrices in the boundary space (i.e. elements in $\mathrm{End}(\mathcal{V}_{B})$) associated to the two latter objects  in the following way
$\mathbf{K}_{i,j}(u)=\left\{ K_{i,k}^{\alpha,\beta}(u)\right\} _{\alpha,\beta=1}^{d_{B}}\in\mathrm{End}(\mathcal{V}_{B})$
and $\langle\Psi|=\left\{ \Psi_{\alpha,\beta}\right\} _{\alpha,\beta=1}^{d_{B}}\in\mathcal{H}^{*}\otimes\mathrm{End}(\mathcal{V}_{B})$.  We can also introduce the notation $\mathbf{K}(u)=\sum_{i,j}e_{i,j}\otimes\mathbf{K}_{i,j}(u)$
for which the KT-relation can be written in a more compact form
\begin{equation}
\mathbf{K}(u)\langle\Psi|T(u)=\langle\Psi|\widehat{T}(-u)\mathbf{K}(u).
\end{equation}
The graphical presentations of the $K$-matrices, boundary states and the KT-relations are shown in figure \ref{fig:KPsi} and \ref{fig:KT}.

\subsection{Compatibility with the twist}

Let us multiply the KT-relation with $G\otimes\mathbf{G}\in GL(4)\otimes GL(d_B)$, i.e.\ we allow for a twist in both the auxiliary and the boundary space
\begin{equation}
\mathbf{K}_{0,B}\langle\Psi_{B}|T_{0}\left(G_{0}\mathbf{G}_{B}\right)=\langle\Psi_{B}|\widehat{T}_{0}\mathbf{K}_{0,B}\left(G_{0}\mathbf{G}_{B}\right),
\end{equation}
where the subscripts denote the vector spaces where the operators
act, and for simplicity we do not write out the spectral parameter
dependence. We can rearrange the l.h.s. and arrive at
\begin{equation}
\mathbf{K}_{0,B}\left(\langle\Psi_{B}|\mathbf{G}_{B}\right)T_{0}G_{0}=\langle\Psi_{B}|\widehat{T}_{0}\mathbf{K}_{0,B}\left(G_{0}\mathbf{G}_{B}\right)\label{eq:KTGG}.
\end{equation}
Let us assume that $G$ is a symmetry of the $K$-matrix, i.e.,
\begin{equation}
\mathbf{K}\left(G\otimes\mathbf{G}\right)=\left(\widehat{G}\otimes\mathbf{G}\right)\mathbf{K}.\label{eq:sym}
\end{equation}
Using this property the equation (\ref{eq:KTGG}) simplifies as 
\[
\mathbf{K}_{0,B}\left(\langle\Psi_{B}|\mathbf{G}_{B}\right)T_{0}G_{0}=\left(\langle\Psi_{B}|\mathbf{G}_{B}\right)\widehat{T}_{0}\widehat{G}_{0}\mathbf{K}_{0,B}.
\]
Assuming $\mathbf{K}$ is invertible, we obtain the crossed integrability
condition
\begin{equation}
\langle \mbox{MPS}|\mathcal{T}(u)=\langle \mbox{MPS}|\widehat{\mathcal{T}}(-u),\label{eq:intCond}
\end{equation}
where we introduced the traced boundary state, i.e.\ the matrix product state 
\begin{equation}
\langle \mbox{MPS}|=\sum_{\alpha,\beta}\langle\psi_{\alpha,\beta}|\mathbf{G}_{\beta,\alpha}=\mathrm{Tr}_{\mathcal{V}_{B}}\left(\langle\Psi|\mathbf{G}\right).
\end{equation}

\subsection{Twisted Yangians}

The compatibility of the crossed KT- and the RTT-relations requires
that the $K$-matrix has to satisfy the crossed reflection equation 
\begin{equation}
R_{1,2}(u-v)\mathbf{K}_{1}(-u)\bar{R}_{1,2}(u+v)\mathbf{K}_{2}(-v)=\mathbf{K}_{2}(-v)\bar{R}_{1,2}(u+v)\mathbf{K}_{1}(-u)R_{1,2}(u-v).
\end{equation}
Let assume that the asymptotic expansion of the $K$-matrix starts as
\begin{equation}
\mathbf{K}(u)=\mathbf{1}+u^{-1}\sum e_{i,j}\otimes F_{i,j}+\mathcal{O}(u^{-2})\label{eq:Femb}.
\end{equation}
The crossed reflection equation with this asymptotic expansion defines
the twisted Yangian $Y^{+}(4)$ algebra. From the reflection equation
we could derive that the operators $F_{i,j}$ generate an $\mathfrak{o}_{4}$
algebra 
\begin{equation}
\begin{split}\left[F_{i,j},F_{k,l}\right] & =\delta_{j,k}F_{i,l}-\delta_{i,l}F_{k,j}-\delta_{j,5-l}F_{i,5-k}+\delta_{i,5-k}F_{5-l,j},\\
F_{5-j,5-i} & =-F_{i,j}.
\end{split}
\end{equation}
This means that the twisted Yangians have $\mathfrak{o}_{4}$ subalgebras.
Therefore any $Y^{+}(4)$ representation (which gives a $K$-matrix) is
an $\mathfrak{o}_{4}$ representation, too.

Let us continue with the symmetry algebra corresponding to the symmetry (\ref{eq:sym})
\begin{equation}
(h\otimes\mathbf{1}+1\otimes\mathbf{h})\mathbf{K}(u)=\mathbf{K}(u)(-h^{t}\otimes\mathbf{1}+1\otimes\mathbf{h}).
\end{equation}
The solution of this equation can be obtained from the series expansion
of the reflection equation \cite{Gombor:2019bun} which leads to
\begin{equation}
h=f_{i,j}=e_{i,j}-e_{5-j,5-i},\quad\mathbf{h}=-F_{j,i},
\end{equation}
for $i,j=1,\dots,4$. Therefore the symmetry algebra is $\mathfrak{o}_{4}$.
From the condition (\ref{eq:sym}) we see that the twist is compatible
with the boundary state only if the twist matrix is orthogonal, i.e.
$G\in SO(4)$. The most general diagonal twist is
\begin{equation}
G=\widehat{G}=\mathrm{diag}(z_{1},z_{2},z_{2}^{-1},z_{1}^{-1}),
\end{equation}
and the corresponding boundary twist matrix is
\begin{equation}
\mathbf{G}=\exp(-\phi_{1}F_{1,1}-\phi_{2}F_{2,2}).
\end{equation}

In the following we will need the highest weights of the twisted Yangian irreps. A vector
$|0\rangle\in\mathcal{V}_{B}$ is highest weight vector if 
\begin{equation}
\begin{split}\mathbf{K}_{i,j}(u)|0\rangle & =0,\quad\text{for }i<j,\\
\mathbf{K}_{i,i}(u)|0\rangle & =\mu_{i}(u)|0\rangle,
\end{split}
\end{equation}
where the functions $\mu_{i}(u)$ are the highest weights. The twisted
Yangian generators satisfy the relation \cite{Molev:1997wp}
\begin{equation}
\mathbf{K}_{5-j,5-i}(-u)=\mathbf{K}_{i,j}(u)+\frac{\mathbf{K}_{i,j}(u)-\mathbf{K}_{i,j}(-u)}{2u},
\end{equation}
therefore $\mu_{1}(u)$ and $\mu_{2}(u)$ are not independent of
$\mu_{3}(u)$ and $\mu_{4}(u)$. 

The highest weights can be used to identify irreps since two irreps
are isomorphic iff the ratios
\begin{equation}
P_{1}(u)=\frac{\mu_{4}(u)}{\mu_{3}(u)},\quad P_{2}(u)=\frac{\mu_{3}(-u)}{\mu_{3}(u)},
\end{equation}
are the same~\cite{Molev:1997wp}.

\subsection{Selection rules for on-shell overlaps \label{sec:onshell}}

We can apply the integrability condition (\ref{eq:intCond}) to an
on-shell Bethe state and get
\begin{equation}
(\tau(v|\bar{u})-\hat{\tau}(-v|\bar{u}))\langle\mathrm{MPS}|\bar{u}\rangle=0
\end{equation}
Non-vanishing on-shell overlaps $\langle\mathrm{MPS}|\bar{u}\rangle\neq0$
require the condition 
\begin{equation}
\tau(v|\bar{u})=\hat{\tau}(-v|\bar{u}).\label{eq:condeig}
\end{equation}
The transfer matrix eigenvalues can be read off from eqn.\ (\ref{eq:eig}) and take the form
\begin{equation}
\begin{split}\tau(v|\bar{u}) & =z_{1}\frac{Q_{\theta}(v+3/2)}{Q_{\theta}(v+1/2)}\frac{Q_{1}(v-1/2)}{Q_{1}(v+1/2)}+z_{2}\frac{Q_{\theta}(v+3/2)}{Q_{\theta}(v+1/2)}\frac{Q_{1}(v+3/2)}{Q_{1}(v+1/2)}\frac{Q_{2}(v)}{Q_{2}(v+1)}+\\
 & +z_{3}\frac{Q_{2}(v+2)}{Q_{2}(v+1)}\frac{Q_{3}(v+1/2)}{Q_{3}(v+3/2)}+z_{4}\frac{Q_{3}(v+5/2)}{Q_{3}(v+3/2)},\\
\hat{\tau}(v|\bar{u}) & =z_{4}^{-1}\frac{Q_{3}(v-5/2)}{Q_{3}(v-3/2)}+z_{3}^{-1}\frac{Q_{2}(v-2)}{Q_{2}(v-1)}\frac{Q_{3}(v-1/2)}{Q_{3}(v-3/2)}+\\
 & +z_{2}^{-1}\frac{Q_{\theta}(v-3/2)}{Q_{\theta}(v-1/2)}\frac{Q_{1}(v-3/2)}{Q_{1}(v-1/2)}\frac{Q_{2}(v)}{Q_{2}(v-1)}+z_{1}^{-1}\frac{Q_{\theta}(v-3/2)}{Q_{\theta}(v-1/2)}\frac{Q_{1}(v+1/2)}{Q_{1}(v-1/2)}.
\end{split}
\end{equation}
For the untwisted case ($z_{j}=1$) we can see that the condition
(\ref{eq:condeig}) is satisfied when
\begin{equation}
Q_{k}(-v)=(-1)^{n_{k}}Q_{k}(v),
\end{equation}
for $k=1,2,3$ which is the well-known pairing condition for the roots~\cite{DeLeeuw:2018cal,Gombor:2020kgu}.

In the twisted case, however, it is not possible to formulate the
selection rule in a nice way with these Bethe roots and $Q$-functions.
To obtain the selection rules in a compact form we need to introduce
the full system of $\mathcal{Q}$-functions \cite{Tsuboi:1998ne} ( $\mathcal{Q}_{j},\mathcal{Q}_{jk},\mathcal{Q}_{jkl}$
where $j,k,l\in\{1,2,3,4\}$),  for reviews see~\cite{Gromov:2019icz,Kazakov:2018ugh,Levkovich-Maslyuk:2019awk}. The 
relevant $\mathcal{Q}\mathcal{Q}$-relations read
\begin{equation}
\begin{split}\mathcal{Q}_{j}(v+1/2)\mathcal{Q}_{k}(v-1/2)-\mathcal{Q}_{j}(v-1/2)\mathcal{Q}_{k}(v+1/2) & \sim\mathcal{Q}_{jk}(v),\\
\mathcal{Q}_{jk}(v+1/2)\mathcal{Q}_{jl}(v-1/2)-\mathcal{Q}_{jk}(v-1/2)\mathcal{Q}_{jl}(v+1/2) & \sim Q_{\theta}(v)\mathcal{Q}_{j}(v)\mathcal{Q}_{jkl}(v),\\
\mathcal{Q}_{jkl}(v+1/2)\mathcal{Q}_{jkn}(v-1/2)-\mathcal{Q}_{jkl}(v-1/2)\mathcal{Q}_{jkn}(v+1/2) & \sim\mathcal{Q}_{jk}(v).
\end{split}
\end{equation}
We can indentify the previous $Q$-functions in terms of the $\mathcal{Q}$-functions as $\mathcal{Q}_{1}(u)=z_{1}^{-u}Q_{1}(u)$,
$\mathcal{Q}_{12}(u)=z_{1}^{-u}z_{2}^{-u}Q_{2}(u)$ and $\mathcal{Q}_{123}(u)=z_{1}^{-u}z_{2}^{-u}z_{3}^{-u}Q_{3}(u)$.
The transfer matrix eigenvalues can be expressed as
\begin{equation}
\begin{split}\tau(v|\bar{u}) & =\frac{Q_{\theta}(v+3/2)}{Q_{\theta}(v+1/2)}\frac{\mathcal{Q}_{j}(v-1/2)}{\mathcal{Q}_{j}(v+1/2)}+\frac{Q_{\theta}(v+3/2)}{Q_{\theta}(v+1/2)}\frac{\mathcal{Q}_{j}(v+3/2)}{\mathcal{Q}_{j}(v+1/2)}\frac{\mathcal{Q}_{jk}(v)}{\mathcal{Q}_{jk}(v+1)}+\\
 & +\frac{\mathcal{Q}_{jk}(v+2)}{\mathcal{Q}_{jk}(v+1)}\frac{\mathcal{Q}_{jkl}(v+1/2)}{\mathcal{Q}_{jkl}(v+3/2)}+\frac{\mathcal{Q}_{jkl}(v+5/2)}{\mathcal{Q}_{jkl}(v+3/2)},\\
\hat{\tau}(v|\bar{u}) & =\frac{\mathcal{Q}_{jkl}(v-5/2)}{\mathcal{Q}_{jkl}(v-3/2)}+\frac{\mathcal{Q}_{jk}(v-2)}{\mathcal{Q}_{jk}(v-1)}\frac{\mathcal{Q}_{jkl}(v-1/2)}{\mathcal{Q}_{jkl}(v-3/2)}+\\
 & +\frac{Q_{\theta}(v-3/2)}{Q_{\theta}(v-1/2)}\frac{\mathcal{Q}_{j}(v-3/2)}{\mathcal{Q}_{j}(v-1/2)}\frac{\mathcal{Q}_{jk}(v)}{\mathcal{Q}_{jk}(v-1)}+\frac{Q_{\theta}(v-3/2)}{Q_{\theta}(v-1/2)}\frac{\mathcal{Q}_{j}(v+1/2)}{\mathcal{Q}_{j}(v-1/2)},
\end{split}
\end{equation}
where we assumed that $z_{1}z_{2}z_{3}z_{4}=1$. We already saw that
the boundary state is compatible with the twist when $z_{j}^{-1}=z_{5-j}$.
We see that the condition (\ref{eq:condeig})  leads to the selection rule
\begin{equation}
\mathcal{Q}_{k}(-v)=\pm\mathcal{Q}_{5-k}(v),
\end{equation}
for $k=1,2,3,4$.

\subsection{Evaluation representations\label{evaluation}}

Evaluation representations are the types of representations which are relevant for the D3-D5 domain wall in
${\cal N}=4$ SYM and are introduced as follows.
 We have an evaluation homomorphism $\mathfrak{o}_{4}\hookrightarrow Y^{+}(4)$:
\begin{equation}\label{evhom}
F_{i,j}\hookrightarrow\mathbf{K}_{i,j}(u)=\delta_{i,j}+\frac{2}{2u+1}F_{i,j},
\end{equation}
therefore we can lift any $\mathfrak{o}_{4}$ irreps to $Y^{+}(4)$
irreps. Since $\mathfrak{o}_{4}\cong\mathfrak{sl}_{2}\oplus\mathfrak{sl}_{2}$,
any 2-tuple $(s_{L},s_{R})$ defines an $\mathfrak{o}_{4}$ highest
weight irrep ($s_{L/R}$ are the spins of the $\mathfrak{sl}_{2}$
irreps). Let us denote by $V(s_{L},s_{R})$ the corresponding $Y^{+}(4)$
irrep.

Furthermore, let us parametrize the $\mathfrak{o}_{4}=\mathfrak{sl}_{2}\oplus\mathfrak{sl}_{2}$
generators as
\begin{align}
F_{1,1} & =S_{3}^{R}+S_{3}^{L}, & F_{2,2} & =S_{3}^{R}-S_{3}^{L},\nonumber \\
F_{1,2} & =\sqrt{2}S_{+}^{L}, & F_{2,1} & =\sqrt{2}S_{-}^{L},\\
F_{1,3} & =\sqrt{2}S_{+}^{R}, & F_{3,1} & =\sqrt{2}S_{-}^{R},\nonumber 
\end{align}
where 
\begin{equation}
\left[S_{j}^{L/R},S_{k}^{L/R}\right]=\sum_{l=1}^{3}i\epsilon_{j,k,l}S_{l}^{L/R},\quad S_{\pm}^{L/R}=\frac{1}{\sqrt{2}}\left(S_{1}^{L/R}\pm iS_{2}^{L/R}\right).
\end{equation}
Using the relation~(\ref{evhom}), the explicit form of the $K$-matrix can easily be written down. We have listed it together
with related quantities of relevance for the present section in appendix~\ref{K-matrices}.

\begin{figure}
\centering
\includegraphics[width=0.35\textwidth]{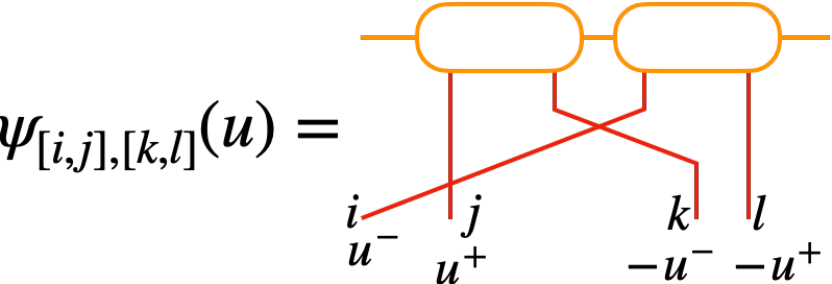} 
\caption{The graphical presentation fusion leading to the boundary state. The red lines correspond to the auxiliary space ($i,j=1,\dots 4$). The boundary space is denoted by orange lines. The intersection of two red lines denote the R-matrix $R(u)$. Here, we used the shorthand notations $u^\pm=u \pm 1/2$ for the spectral parameter dependence. The rounded rectangles are the two-site boundary operators $\psi^{(0)}_{i,j}(u)=\mathbf{K}_{5-j,i}(u)$ for the defining representation of $\mathfrak{gl}_4$ .}
\label{fig:fusion}
\end{figure}

The corresponding boundary state can be obtained using the fusion procedure sketched in figure \ref{fig:fusion}. If the quantum space were the defining representation the boundary state would be constructed from the two-site operators $\psi^{(0)}_{i,j}(u)=\mathbf{K}_{5-j,i}(u)$. The fused two-site operator is
\begin{equation}
\begin{split}
\psi_{[i,j],[k,l]}(u) =& -u^2 (\psi^{(0)}_{i,l}(u-1/2) \psi^{(0)}_{j,k}(u+1/2) - \psi^{(0)}_{j,l}(u-1/2) \psi^{(0)}_{i,k}(u+1/2)) - \\
       & -\frac{u}{2} (\psi^{(0)}_{k,l}(u-1/2) \psi^{(0)}_{j,i}(u+1/2) - \psi^{(0)}_{k,l}(u-1/2) \psi^{(0)}_{i,j}(u+1/2)).
\end{split}
\end{equation}
The explicit form of the fused two-site operator $\psi_{a,b}(u)$, $a,b=1,\ldots 6$, is given in appendix~\ref{K-matrices}. From
this two-site operator one builds the boundary state as
\begin{equation}
\langle\Psi|=\sum_{i_{1},\dots,i_{2J}}\langle i_{1},i_{2},\dots,i_{2J-1},i_{2J}|\otimes\psi_{i_{2J-1},i_{2J}}(\theta_{J})\dots\psi_{i_{1},i_{2}}(\theta_{1}).\label{eq:BS}
\end{equation}
The corresponding MPS can be written as
\begin{equation}
\langle \mbox{MPS}|=\sum_{i_{1},\dots,i_{2J}}\mathrm{Tr}\left[\psi_{i_{2J-1},i_{2J}}(\theta_{J})\dots\psi_{i_{1},i_{2}}(\theta_{1})\mathbf{G}\right]\langle i_{1},i_{2},\dots,i_{2J-1},i_{2J}|,\label{eq:MPS}
\end{equation}
where the boundary twist matrix is defined as
\begin{equation}
\mathbf{G}=\exp(-\phi_{1}(S_{3}^{R}+S_{3}^{L})-\phi_{2}(S_{3}^{R}-S_{3}^{L}))=\exp((\phi_{2}-\phi_{1})S_{3}^{L}-(\phi_{1}+\phi_{2})S_{3}^{R}).
\end{equation}
For $u=0$ the elementary building block $\psi_{a,b}(u)$ of the MPS
is factorized as~(cf.\ appendix~\ref{K-matrices})
\begin{equation}
\psi_{a,b}(0)=\omega_{b}\omega_{a}, 
\end{equation}
where
\begin{equation}
\bar{\omega}  =\left\{ \omega_{Z},\omega_{Y},\omega_{X},\omega_{\bar{X}},\omega_{\bar{Y}},\omega_{\bar{Z}}\right\} =\left\{ \sqrt{2}S_{-}^{R},\sqrt{2}S_{-}^{L},S_{3}^{R}+S_{3}^{L},S_{3}^{R}-S_{3}^{L},-\sqrt{2}S_{+}^{L},\sqrt{2}S_{+}^{R}\right\}.
\end{equation}
In the homogenous limit the MPS simplifies as
\begin{equation}
\langle \mbox{MPS}|=\sum_{i_{1},\dots,i_{2J}}\mathrm{Tr}\left[\omega_{i_{2J}}\dots\omega_{i_{2}}\omega_{i_{1}}\mathbf{G}\right]\langle i_{1},i_{2},\dots,i_{2J}|.
\end{equation}
We define two special classes of matrix product states. For the first one we set $s_{L}=0$
and $s_{R}=s$ and denote the associated MPS as  $\langle \mbox{MPS}_{2s+1}^{R}|$. We give the associated
$K$-matrix, $\mathbf{K}^R(u)$ in appendix~\ref{K-matrices}.
 The highest weights of the corresponding $Y^+(4)$ representation are
\begin{align}
\mu_{4}^{R}(u)  =\frac{(u+1/2-s)}{(u+1/2)}, \hspace{0.5cm}
\mu_{3}^{R}(u)  =\frac{(u+1/2-s)}{(u+1/2)},
\end{align}
therefore
\begin{equation}
P_{1}^{R}(u)=1,\quad P_{2}^{R}(u)=\frac{u+1/2}{u-1/2}\frac{u-1/2+s}{u+1/2-s}.\label{eq:ratio}
\end{equation}
We will denote this representation as $V_{R}(s)\equiv V(0,s)$.

We also define the special representations where $s_{R}=0$ and $s_{L}=s$,
for which we denote the associated MPS as $\langle \mbox{MPS}_{2s+1}^{L}|$. Again we give the explicit
form of the corresponding $K$-matrix, $\mathbf{K}^L(u)$ in appendix~\ref{K-matrices}.
The highest weights of the corresponding $Y^+(4)$ representation are
\begin{equation}
\mu_{4}^{L}(u)=\frac{u+1/2-s}{u+1/2},\quad\mu_{3}^{L}(u)=\frac{u+1/2+s}{u+1/2}.
\end{equation}
Therefore
\begin{equation}
P_{1}^{L}(u)=\frac{u+1/2-s}{u+1/2+s},\quad P_{2}^{L}(u)=\frac{u+1/2}{u-1/2}\frac{u-1/2-s}{u+1/2+s}.\label{eq:ratio-1}
\end{equation}
We will denote this representation as $V_{L}(s)\equiv V(s,0)$.  The representations which are of relevance for the D3-D5 domain wall problem are the representations of type $V_{L}(s)$ and $V_{R}(s)$, cf.~section~\ref{D3D5}.

\section{Dressing\label{Main_Dressing}}
Let us assume a $K$-matrix which solves the KT relation is known. Starting from this solution one can can obtain another solution be means of quantum dressing as we will now explain.
The twisted Yangian $Y^{+}(4)$ has the following co-module property
\cite{Molev:1997wp}:
\begin{equation}
\Delta:Y^{+}(4)\to Y(4)\otimes Y^{+}(4),
\end{equation}
which is defined as
\begin{equation}
\Delta(\mathbf{K}_{i,j}(u))=\sum_{k,l=1}^{4}T_{5-k,5-i}(u)T_{l,j}(-u)\otimes\mathbf{K}_{k,l}(u).
\end{equation}
This can be used to ''dress'' representations of the twisted Yangians
$Y^{+}(4)$ with the original Yangian $Y(4)$. For the original Yangian
the evaluation homomorphism is
\begin{equation}
E_{i,j}\hookrightarrow T_{i,j}(u)=\delta_{i,j}+\frac{1}{u}E_{j,i},
\end{equation}
where $E_{i,j}$ are the $\mathfrak{gl}_{4}$ generators. For the
$4$-tuples $\lambda\equiv(\lambda_{1},\lambda_{2},\lambda_{3},\lambda_{4})$
we have the highest weight $\mathfrak{gl}_{4}$ representations $E_{i,j}^{\lambda}$ and the
corresponding $Y(4)$ representations, $L(\lambda|\xi)$, which can be identified with the
quantities denoted as $\mathcal{L}^{\lambda,\mathbf{4}}(u-\xi)$ in section~\ref{notation},
where $\xi$ was  an inhomogeneity. We can always choose $\lambda_{4}=0$
since
\begin{equation}
L(\lambda_{1},\dots,\lambda_{4}|\xi)\cong L(\lambda_{1}-\xi,\dots,\lambda_{4}-\xi|0).
\end{equation}
Let us pick a $Y^{+}(4)$ representation $\mathcal{V}_{B}=V$. After
the dressing with $L(\lambda|\xi)$ we have a $Y^{+}(4)$ representation
$\mathcal{V}_{B}=L(\lambda|\xi)\otimes V$ which can be written as
\begin{equation}
\widetilde{\mathbf{K}}_{0,L,V}(u)=\left[\mathcal{L}_{L,0}^{\lambda,\mathbf{4}}(u-\xi)\right]^{t_{0}}\mathbf{K}_{0,V}(u)\,\mathcal{L}_{L,0}^{\lambda,\mathbf{4}}(-u-\xi).\label{eq:dressed0}
\end{equation}
where the subscripts 0, $L$ and $V$ denote the vector spaces where
the operators act. We can introduce the notation
\begin{equation}
\bar{\mathcal{L}}_{L,0}^{\lambda,\mathbf{4}}(u)\equiv \left[\mathcal{L}_{L,0}^{\lambda,\mathbf{4}}(-u)\right]^{t_{0}},
\end{equation}
for which the dressed K-matrix (\ref{eq:dressed0}) can be written
in a more compact form
\begin{equation}
\widetilde{\mathbf{K}}_{0,L,V}(u)=\bar{\mathcal{L}}_{L,0}^{\lambda,\mathbf{4}}(\xi-u)\mathbf{K}_{0,V}(u)\mathcal{L}_{L,0}^{\lambda,\mathbf{4}}(-u-\xi).
\end{equation}
We can check that this matrix satisfies the reflection equation
\begin{equation}
R_{1,2}(u-v)\widetilde{\mathbf{K}}_{1,L,V}(-u)\bar{R}_{1,2}(u+v)\widetilde{\mathbf{K}}_{2,L,V}(-v)=\widetilde{\mathbf{K}}_{2,L,V}(-v)\bar{R}_{1,2}(u+v)\widetilde{\mathbf{K}}_{1,L,V}(-u)R_{1,2}(u-v),
\end{equation}
using the bYB for the original $\mathbf{K}_{0,V}$ and the
RLL-relations~(\ref{RLL}).

The natural question is: what is the integrable boundary state corresponding to the dressed $K$-matrix, cf.\ eqn.~(\ref{KT}). 
Let
us introduce the dressed boundary state 
\begin{equation}
\langle\tilde{\Psi}_{L,V}|=\langle\Psi_{V}|T_{L}^{\lambda}(-\xi),
\end{equation}
where we used the monodromy matrix where the representation of the
auxiliary space is $\lambda$. It is straightforward to show that this dressed boundary
state $\langle\tilde{\Psi}_{L,V}|$ satisfies the KT-relation with
the dressed K-matrix:
\begin{equation}
\widetilde{\mathbf{K}}_{0,L,V}(z)\langle\tilde{\Psi}_{L,V}|T_{0}(z)=\langle\tilde{\Psi}_{L,V}|\widehat{T}_{0}(-z)\widetilde{\mathbf{K}}_{0,L,V}(z).
\end{equation}
One only needs the KT-relation for the original state $\langle\Psi_{V}|$
and the RTT-relations~(\ref{eq:comm1}).

We can apply the dressing procedure to the one-dimensional representation, i.e.\
when the $K$-matrix is
\begin{equation}
\mathbf{K}_{i,j}=\delta_{i,j},
\end{equation}
and the boundary state is the simple two-site state
\begin{equation}
\langle\delta|=\langle\varphi|^{\otimes J},\quad\langle\varphi|=\sum_{a,b}\varphi_{a,b}\langle a,b|\label{eq:deltastate},
\end{equation}
where $\varphi_{i,j}$ is given in eqn.~(\ref{phiij}).
After the dressing  the representation is $\mathcal{V}_{B}=L(\lambda|\xi)$
with the following dressed $K$-matrix and associated integrable boundary state
\begin{equation}
\mathbf{K}_{0,L}(z)=\bar{\mathcal{L}}_{L,0}^{\lambda,\mathbf{4}}(\xi-z)\mathcal{L}_{L,0}^{\lambda,\mathbf{4}}(-z-\xi),\quad\langle\Psi_{L}|=\langle\delta|T_{L}^{\lambda}(-\xi).\label{eq:dressed}
\end{equation}
Taking the trace we get a twisted MPS which is simply the result of  acting with a transfer matrix on the basic two-site
state
\begin{equation}
\langle \mbox{MPS}|=\langle\delta|\mathcal{T}_{\lambda}(-\xi)\label{eq:dressedStates}.
\end{equation}

We can check that the lowest weight state of the $\mathfrak{gl}_{4}$
is the highest weight state of the dressed $K$-matrix. The highest weights
of the dressed representation $L(\lambda|\xi)$ are
\begin{equation}
\mu_{4}^{D}(u)=\frac{(u+\xi-\lambda_{1})(u-\xi+\lambda_{4})}{(u+\xi)(u-\xi)},\quad\mu_{3}^{D}(u)=\frac{(u+\xi-\lambda_{2})(u-\xi+\lambda_{3})}{(u+\xi)(u-\xi)},
\end{equation}
therefore
\begin{equation}
P_{1}^{D}(u)=\frac{(u+\xi-\lambda_{1})(u-\xi+\lambda_{4})}{(u+\xi-\lambda_{2})(u-\xi+\lambda_{3})},\quad P_{2}^{D}(u)=\frac{(u-\xi+\lambda_{2})(u+\xi-\lambda_{3})}{(u+\xi-\lambda_{2})(u-\xi+\lambda_{3})}.\label{eq:dressedP}
\end{equation}
Let us fix $\lambda_{4}=0$. The ratios of highest weights are same
as $V_{R}(s)$ (\ref{eq:ratio}), if $\lambda_{1}=\lambda_{2}=s$,
$\lambda_{3}=\lambda_{4}=0$ and $\xi=1/2$.

The advantage of the dressed representations is that the overlap has
been reduced to the two-site state overlaps, since:

\begin{equation}
\frac{\langle \mbox{MPS}|\bar{u}\rangle}{\langle\delta|\bar{u}\rangle}=\tau_{\lambda}(-\xi).
\end{equation}

\subsection{Generalized dressing formulas for MPSs\label{gen_dressing}}

In the previous section we saw that for special MPSs obtained by dressing the 
K-matrix corresponding to the one-dimensional representation of $Y^+(4)$ by $L(\lambda|\xi)$
there exists a simple, closed formula for the overlap with Bethe eigenstates, because
the MPS can be expressed as an action of a transfer matrix on the simple two site product state $|\delta\rangle$. 

For general representations of $Y^+(4)$ such a simple relation to the state $|\delta\rangle$
does not exist. However, it is possible to express the general integrable MPS by a
linear combination of the action of transfer matrices on this state, i.e.\
\begin{equation}
\langle \mbox{MPS}|=\langle\delta|\left[\sum_{j=1}^{k}F_{j}\mathcal{T}_{\lambda^{(j)}}(-\xi^{(j)})\right],
\end{equation}
where $F_{j}\in\mathbb{C}$. 
Thus, in the general case the overlap quotient is a linear combination of transfer matrix eigenvalues
\begin{equation}
\frac{\langle \mbox{MPS}|\bar{u}\rangle}{\langle\delta|\bar{u}\rangle}=\sum_{j=1}^{k}F_{j}\,\tau_{\lambda^{(j)}}(-\xi^{(j)}),
\end{equation}
i.e.\ the MPS overlap problem is reduced to the much simpler overlap problem for the basic two-site product state $|\delta\rangle$.
 For untwisted Bethe vectors this problem was already solved \cite{Gombor:2021hmj,Gombor:2023bez}.
For twisted Bethe states the functional
SoV approach might provide a way forward \cite{Sklyanin:1995bm,Gromov:2016itr,Ekhammar:2023iph}. 

Our goal is to find the generalized dressing formulas and the overlap ratios for the states $\langle \mbox{MPS}_{2s+1}^{L/R}|$ which will allow
us to fill the last gap in the proof of the expression for the tree level one-point functions of scalars in 
the D3-D5 domain wall version of ${\cal N}=4$ SYM.
We follow the strategy of \cite{DeLeeuw:2019ohp} where the overlap ratio
was determined for simpler MPSs describing the tree-level one-point functions in the $SU(3)$ sub-sector. In the first subsection we review
the method of \cite{DeLeeuw:2019ohp}, next we apply it to
the $\langle \mbox{MPS}_{2s+1}^{L/R}|$.

\subsection{Methodology \label{recursion}}

In the following we describe our methodology with a general
example. Let us pick a $Y^{+}(4)$ irrep $V\equiv V^{(1)}$ with the
corresponding $K$-matrix $\mathbf{K}_{i,j}(u)\equiv\mathbf{K}_{i,j}^{(1)}(u)$
and boundary state $\langle\Psi|\equiv\langle\Psi^{(1)}|$ which satisfy
the KT-relation. The matrix product state $\langle \mbox{MPS}|\equiv\langle \mbox{MPS}^{(1)}|$ is
defined from the boundary state by taking the twisted trace in the
boundary space.

At first, we need to find a $Y(4)$ representation $L(\lambda|\xi)\equiv L^{(1)}(\lambda^{(1)}|\xi^{(1)})$
for which the dressed K-matrix $\mathbf{K}_{i,j}^{D}(u)\equiv\mathbf{K}_{i,j}^{D_{1}}(u)$
(given by (\ref{eq:dressed})) has the same ratio of highest
weights $P_{k}(u)$ as $\mathbf{K}_{i,j}^{(1)}(u)$. 
If the representation $L(\lambda|\xi)\equiv L^{(1)}(\lambda^{(1)}|\xi^{(1)})$
has the same dimension as $V^{(1)}$ then the two representations
are isomorphic, i.e., $L^{(1)}(\lambda^{(1)}|\xi^{(1)})=V^{(1)}$.
Choosing a proper basis in the boundary space the two $K$-matrices
are the same $\mathbf{K}_{i,j}^{D_{1}}(u)=\mathbf{K}_{i,j}^{(1)}(u)$, and 
therefore the boundary states are also the same
\begin{equation}
\langle\Psi^{D_{1}}|=\langle\delta|T^{\lambda^{(1)}}(-\xi^{(1)})=\langle\Psi^{(1)}|.
\end{equation}
Taking the twisted trace in the boundary space we obtain that 
\begin{equation}
\langle \mbox{MPS}^{(1)}|=\langle\delta|\mathcal{T}_{\lambda^{(1)}}(-\xi^{(1)}),
\end{equation}
therefore the ratio of the overlaps is simply a transfer matrix eigenvalue
\begin{equation}
\frac{\langle \mbox{MPS}^{(1)}|\bar{u}\rangle}{\langle\delta|\bar{u}\rangle}=\tau_{\lambda^{(1)}}(-\xi^{(1)}).
\end{equation}

Let us continue with the other case when the dimension of $L^{(1)}(\lambda^{(1)}|\xi^{(1)})$
is bigger than the dimension of $V^{(1)}$. We know that  two irreps with the same $P_{k}(u)$'s
are isomorphic, therefore the representation $L^{(1)}(\lambda^{(1)}|\xi^{(1)})$
can not be irreducible, and it contains a $V^{(1)}$ irrep, i.e., we
have the decomposition
\begin{equation}
L^{(1)}(\lambda^{(1)}|\xi^{(1)})=V^{(1)}\oplus V^{(2)}.\label{eq:decomp}
\end{equation}
This decomposition is not necessarily a direct sum, it can be semi-direct
sum. It means that the dressed $K$-matrix has the following block
form
\begin{equation}
\mathbf{K}_{i,j}^{D_{1}}(u)=\left(\begin{array}{cc}
\mathbf{K}_{i,j}^{(1)}(u) & X_{i,j}^{(1)}\\
0 & \mathbf{K}_{i,j}^{(2)}(u)
\end{array}\right),\quad\text{or}\quad\mathbf{K}_{i,j}^{D_{1}}(u)=\left(\begin{array}{cc}
\mathbf{K}_{i,j}^{(1)}(u) & 0\\
X_{i,j}^{(1)} & \mathbf{K}_{i,j}^{(2)}(u)
\end{array}\right),\label{eq:Kdecomp}
\end{equation}
where $\mathbf{K}_{i,j}^{(2)}(u)$ is the K-matrix of the representation
$V^{(2)}$. For these K-matrices there are corresponding boundary states, and it follows that 
these should have the same block structure in the boundary space.
\begin{equation}
\langle\Psi^{D_{1}}|=\left(\begin{array}{cc}
\langle\Psi^{(1)}| & \langle X|\\
0 & \langle\Psi^{(2)}|
\end{array}\right),\quad\text{or}\quad\langle\Psi^{D_{1}}|=\left(\begin{array}{cc}
\langle\Psi^{(1)}| & 0\\
\langle X| & \langle\Psi^{(2)}|
\end{array}\right).
\end{equation}
Taking the (twisted) trace in the boundary space we obtain the following relation
\begin{equation}
\langle \mbox{MPS}^{(1)}|=\langle\delta|\mathcal{T}_{\lambda^{(1)}}(-\xi^{(1)})-\langle \mbox{MPS}^{(2)}|.
\end{equation}
We have thus reduced the original MPS to a linear combination of the dressed
state and a new MPS.  Next, we repeat the steps above starting with 
$V^{(2)}, \mathbf{K}_{i,j}^{(2)}(u)$ and $\langle \mbox{MPS}^{(2)}|$. If $V^{(2)}$ is (isomorphic to) an irrep of $Y^+(4)$ we
are basically done. If not we must continue the process above to obtain yet another representation 
 $V^{(3)},\mathbf{K}_{i,j}^{(3)}(u),\langle \mbox{MPS}^{(3)}|$. Assuming that the process stops after 
$k$ steps, i.e.
i.e. $V^{(k)}=L^{(k)}(\lambda^{(k)}|\xi^{(k)})$
the original MPS can be expressed as
\begin{equation}
\langle \mbox{MPS}^{(1)}|=F_{1}\langle\delta|\mathcal{T}_{\lambda^{(1)}}(-\xi^{(1)})+F_{2}\langle\delta|\mathcal{T}_{\lambda^{(2)}}(-\xi^{(2)})+\dots+F_{k}\langle\delta|\mathcal{T}_{\lambda^{(k)}}(-\xi^{(k)}).
\end{equation}
Since the $KT$-relation is linear in both the $K$-matrix and the boundary
state the normalizations are not fixed, and therefore we have to introduce the
scalars $F_{j}$. In sections~\ref{dressing}
and appendices~\ref{integerbranching} and \ref{halfintegerbranching} we will show how the recursive procedure
works in practice. 

\subsection{Generalized dressing for fixed spin \label{dressing}}

In this subsections we will analyze  decompositions of the type (\ref{eq:decomp}) for fixed spin.
We use the $Y^{+}(4)\to\mathfrak{o}_{4}$ embedding (\ref{eq:Femb}),
i.e. the $\mathfrak{o}_{4}$ subalgebra of the twisted Yangian. The
dressed K-matrix of the representation $L(\lambda|\xi)$ can be series
expanded as 
\begin{equation}
\mathbf{K}_{i,j}^{\lambda}(u)=\mathcal{L}_{5-k,5-i}^{\lambda,\mathbf{4}}(u-\xi)\mathcal{L}_{k,j}^{\lambda,\mathbf{4}}(-u-\xi)=\delta_{i,j}+u^{-1}\left(E_{5-i,5-j}^{\lambda}-E_{j,i}^{\lambda}\right)+\mathcal{O}(u^{-2}),
\end{equation}
which follows from the definition of the $\mathcal{L}$'s, cf.~section~\ref{notation}. 
Since the $u^{-1}$ term defines the $\mathfrak{o}_{4}$ subalgebra
and the dressed $K$-matrices are defined from a $\mathfrak{gl}_{4}$
representation we also have another algebra embedding $\mathfrak{gl}_{4}\to\mathfrak{o}_{4}$
where the $\mathfrak{o}_{4}$ subalgebra of $\mathfrak{gl}_{4}$ is
defined as
\begin{equation}
F_{i,j}=E_{5-i,5-j}-E_{j,i}.
\end{equation}
The states $|\Lambda\rangle$ in $L(\lambda|\xi)$ have $\mathfrak{gl}_{4}$
weights $(\omega_{1},\omega_{2},\omega_{3},\omega_{4})$ but they
also have $\mathfrak{o}_{4}$ spins $(s_{L},s_{R})$ where these are
defined as
\begin{equation}
\begin{split}S_{3}^{R} & =\frac{1}{2}\left(F_{1,1}+F_{2,2}\right)=\frac{1}{2}\left(-E_{1,1}-E_{2,2}+E_{3,3}+E_{4,4}\right),\\
S_{3}^{L} & =\frac{1}{2}\left(F_{1,1}-F_{2,2}\right)=\frac{1}{2}\left(-E_{1,1}+E_{2,2}-E_{3,3}+E_{4,4}\right),
\end{split}
\end{equation}
i.e. the $\mathfrak{o}_{4}$ spins $(s_{L},s_{R})$ of the state $|\Lambda\rangle$
are
\begin{equation}
\begin{split}s_{L} & =\frac{1}{2}\left(-\omega_{1}+\omega_{2}-\omega_{3}+\omega_{4}\right),\\
s_{R} & =\frac{1}{2}\left(-\omega_{1}-\omega_{2}+\omega_{3}+\omega_{4}\right).
\end{split}
\end{equation}
\paragraph{Integer spins}
We now apply the methodology of the previous section to the case of evaluation representations corresponding to integer
spin. They key point is to realize that the decomposition of the $L((s,s,0,0)|-1/2)$ representation of $Y^+(4)$ 
into irreps always contains the the direct sum $V_L(s) \oplus V_R(s)$ and that the remaining components can be found
in closed form as well. In appendix~\ref{integerbranching} we show how the decomposition works for $s=1$, $s=2$ and $s=3$ and present
the general branching rule  in eqn.~(\ref{eq:branching}).
Using this rule we obtain the generalized
dressing formula for integer spins
\begin{multline}
\langle \mbox{MPS}_{2s+1}^{R}|+\langle \mbox{MPS}_{2s+1}^{L}|=\langle\delta|\mathcal{T}_{(s,s,0,0)}(-1/2)+\\
+\frac{Q_{\theta}(s)}{Q_{\theta}(1)}\langle\delta|\left(\mathcal{T}_{(s-1,s-3,0,0)}(-s+1/2)-\mathcal{T}_{(s,s-2,0,0)}(-s+1/2)-\mathcal{T}_{(s-3,s-3,0,0)}(-s+1/2)\right),\label{eq:identity}
\end{multline}
where the normalization factor was fitted numerically.

\paragraph{Half integer spins \label{halfintegerdressing}}

For half integer spins we cannot follow the previous strategy since
there is no finite dimensional $\mathfrak{gl}_{4}$ irrep for which
the ratio of highest weights (\ref{eq:dressedP}) is the same as (\ref{eq:ratio})
for half integer spins. This suggests that the MPSs for half integer
spins cannot be expressed through a two-site state. Therefore, in this
case we look for another type of dressing formula. Our goal will be to relate
the general half integer spin states $\langle \mbox{MPS}_{2s+1}^{R}|$ 
to the 1/2 spin states $\langle\sigma^{R}|\equiv\langle \mbox{MPS}_{2}^{R}|$
and $\langle\sigma^{L}|\equiv\langle \mbox{MPS}_{2}^{L}|$.

In the following we denote the half integers as $s+1/2$ where $s$
is integer. Using this convention, the highest weights of $V_{R}(s+1/2)$
is
\begin{equation}
P_{1}^{R}(u)=1,\quad P_{2}^{R}(u)=\frac{(u+1/2)(u+s)}{(u-1/2)(u-s)}.\label{eq:Phalfint}
\end{equation}
Now we dress the $V_{R}(1/2)$ and $V_{L}(1/2)$ K-matrices and boundary
states. Therefore the dressed K-matrix is
\begin{equation}
\mathbf{K}_{i,j}^{D}(u)=\sum_{k,l=1}^{4}\mathcal{L}_{5-k,5-i}^{\lambda,\mathbf{4}}(\xi-u)\mathcal{L}_{l,j}^{\lambda,\mathbf{4}}(-u-\xi)\otimes\mathbf{K}_{k,l}^{L/R}(u)\in\mathrm{End}\left(L(\lambda|\xi)\otimes V_{L/R}(1/2)\right),\label{eq:halfintK}
\end{equation}
where $\mathbf{K}^{L/R}(u)$ are the K-matrices of the representations
$V_{L/R}(1/2)$. We will denote the basis of the two dimensional space
$V_{L/R}(1/2)$ as $|+\rangle$ and $|-\rangle$ for which
\begin{equation}
S_{3}^{L/R}|\pm\rangle=\pm\frac{1}{2}|\pm\rangle,\quad S_{\pm}^{L/R}|\pm\rangle=0,\quad S_{\mp}^{L/R}|\pm\rangle=\frac{1}{\sqrt{2}}|\mp\rangle.
\end{equation}
One can calculate the highest weights of $L(\lambda|\xi)\otimes V_{R}(1/2)$
(for which the K-matrix is given by (\ref{eq:halfintK})):
\begin{equation}
\begin{split}\mu_{4}(u) & =\frac{u}{u+1/2}\frac{(u+\xi-\lambda_{1})(u-\xi+\lambda_{4})}{(u+\xi)(u-\xi)},\quad\mu_{3}(u)=\frac{u}{u+1/2}\frac{(u+\xi-\lambda_{2})(u-\xi+\lambda_{3})}{(u+\xi)(u-\xi)},\\
P_{1}(u) & =\frac{(u+\xi-\lambda_{1})(u-\xi+\lambda_{4})}{(u+\xi-\lambda_{2})(u-\xi+\lambda_{3})},\quad P_{2}(u)=\frac{u+1/2}{u-1/2}\frac{(u-\xi+\lambda_{2})(u+\xi-\lambda_{3})}{(u+\xi-\lambda_{2})(u-\xi+\lambda_{3})},
\end{split}
\end{equation}
which are the same as (\ref{eq:Phalfint}) when $\lambda_{3}=\lambda_{4}=0$,
$\lambda_{1}=\lambda_{2}=s$ and $\xi=0$. This means that the representation $L((s,s,0,0)|0)\otimes V_{R}(1/2)$
is not irreducible , i.e.,
\begin{equation}
L((s,s,0,0)|0)\otimes V_{R}(1/2)=V_{R}(s+1/2)\oplus V^{(2)}. \label{firststep}
\end{equation}
We also need highest weights of $L(\lambda|\xi)\otimes V_{L}(1/2)$:
\begin{equation}
\begin{split}\mu_{4}(u) & =\frac{u}{u+1/2}\frac{(u+\xi-\lambda_{1})(u-\xi+\lambda_{4})}{(u+\xi)(u-\xi)},\quad\mu_{3}(u)=\frac{u+1}{u+1/2}\frac{(u+\xi-\lambda_{2})(u-\xi+\lambda_{3})}{(u+\xi)(u-\xi)},\\
P_{1}(u) & =\frac{u}{u+1}\frac{(u+\xi-\lambda_{1})(u-\xi+\lambda_{4})}{(u+\xi-\lambda_{2})(u-\xi+\lambda_{3})},\quad P_{2}(u)=\frac{(u-1)(u+1/2)}{(u+1)(u-1/2)}\frac{(u-\xi+\lambda_{2})(u+\xi-\lambda_{3})}{(u+\xi-\lambda_{2})(u-\xi+\lambda_{3})}.
\end{split}
\end{equation}

We now apply the methodology of section~\ref{recursion} to relate $V^{(2)}$ to irreducible representations.
  In appendix~\ref{halfintegerbranching} we show how the decomposition works for $s=1$ and $s=2$ and  present the general branching rule  in eqn.~(\ref{eq:branching-1}).
Using this rule we obtain the generalized
dressing formula for half integer spins
\begin{multline}
\langle \mbox{MPS}_{2s+2}^{R}|=\frac{1}{Q_{\theta}(1/2)}\langle\sigma^{R}|\mathcal{T}_{(s,s,0,0)}(0)+\\
+\frac{Q_{\theta}(s+1/2)}{Q_{\theta}(1/2)^{2}}\left(\langle\sigma^{L}|\mathcal{T}_{(s-1,s-2,0,0)}(-s)-\langle\sigma^{L}|\mathcal{T}_{(s,s-1,0,0)}(-s)-\langle\sigma^{R}|\mathcal{T}_{(s-2,s-2,0,0)}(-s)\right),\label{eq:halfintdressing}
\end{multline}
where the normalization factor was fitted numerically.

\subsection{Universality of the generalized dressing formulas wrt.\ the quantum space}

Since our calculation is based on the $K$-matrix (which is universal,
i.e., do not depend on the representation of the quantum space) the
generalized dressing formulas (\ref{eq:identity}) and (\ref{eq:halfintdressing})
are true for any representations of the quantum space (up to the normalization
scalars). To demonstrate this universality let us pick another example.
Let us change the quantum space to defining representation of $\mathfrak{gl}_{4}$
(usual $SU(4)$ spin chain). Since the algebra is the same (RTT and
the KT are the same), we only need to modify the explicit form the
Lax-operators 
\begin{align}
\mathcal{L}^{\mathbf{4},\mathbf{4}}(u) & =\sum_{i,j=1}^{4}e_{i,j}\otimes\left(\delta_{i,j}+\frac{1}{u}e_{j,i}\right),\qquad\widehat{\mathcal{L}}^{\mathbf{4},\mathbf{4}}(u)=\sum_{i,j=1}^{4}e_{5-i,5-j}\otimes\left(\delta_{i,j}+\frac{1}{-u}e_{i,j}\right),\\
\mathcal{L}^{\lambda,\mathbf{4}}(u) & =\sum_{i,j=1}^{4}\left(\delta_{i,j}+\frac{1}{u}E_{j,i}^{\lambda}\right)\otimes e_{i,j},
\end{align}
which gives the monodromy matrices 
\begin{align}
T_{0}(u) & =\mathcal{L}_{0,2J}^{\mathbf{4},\mathbf{4}}(u+\theta_{J})\mathcal{L}_{0,2J-1}^{\mathbf{4},\mathbf{4}}(u-\theta_{J})\dots\mathcal{L}_{0,2}^{\mathbf{4},\mathbf{4}}(u+\theta_{1})\mathcal{L}_{0,1}^{\mathbf{4},\mathbf{4}}(u-\theta_{1}),\nonumber \\
\widehat{T}_{0}(u) & =\widehat{\mathcal{L}}_{0,2J}^{\mathbf{4},\mathbf{4}}(u+\theta_{J})\widehat{\mathcal{L}}_{0,2J-1}^{\mathbf{4},\mathbf{4}}(u-\theta_{J})\dots\widehat{\mathcal{L}}_{0,2}^{\mathbf{4},\mathbf{4}}(u+\theta_{1})\widehat{\mathcal{L}}_{0,1}^{\mathbf{4},\mathbf{4}}(u-\theta_{1}),\\
T_{0}^{\lambda}(u) & =\mathcal{L}_{0,2J}^{\lambda,\mathbf{4}}(u+\theta_{J})\mathcal{L}_{0,2J-1}^{\lambda,\mathbf{4}}(u-\theta_{J})\dots\mathcal{L}_{0,2}^{\lambda,\mathbf{4}}(u+\theta_{1})\mathcal{L}_{0,1}^{\lambda,\mathbf{4}}(u-\theta_{1}).\nonumber 
\end{align}
After this, the algebra of the monodromy matrix, the definition of
the transfer matrices and the twists are the same. The forms of the
boundary state and MPS are also the same as before, i.e.\ (\ref{eq:BS}), (\ref{eq:MPS}) still hold, and
we only need to update $\psi_{i,j}(u)$ as
\begin{equation}
\psi_{i,j}(u)=\mathbf{K}_{5-j,i}(u).
\end{equation}
The MPS for integer spins can be expressed in the same way as before
\begin{align}
 & \langle \mbox{MPS}_{2s+1}^{R}|+\langle \mbox{MPS}_{2s+1}^{L}|=\langle\delta|\mathcal{T}_{(s,s,0,0)}(-1/2)+\label{eq:identity-1}\\
 & +\frac{Q_{\theta}(s-1/2)}{Q_{\theta}(1/2)}\langle\delta|\left(\mathcal{T}_{(s-1,s-3,0,0)}(-s+1/2)-\mathcal{T}_{(s,s-2,0,0)}(-s+1/2)-\mathcal{T}_{(s-3,s-3,0,0)}(-s+1/2)\right), \nonumber 
\end{align}
where we only needed to update the normalization factors. The MPS for half integer spins can similarly be expressed as 
\begin{multline}
\langle \mbox{MPS}_{2s+2}^{R}|=\langle\sigma^{R}|\mathcal{T}_{(s,s,0,0)}(0)+\\
+\frac{Q_{\theta}(s)}{Q_{\theta}(0)}\left(\langle\sigma^{L}|\mathcal{T}_{(s-1,s-2,0,0)}(-s)-\langle\sigma^{L}|\mathcal{T}_{(s,s-1,0,0)}(-s)-\langle\sigma^{R}|\mathcal{T}_{(s-2,s-2,0,0)}(-s)\right), \label{subtle}
\end{multline}
Note that when the taking the homogeneous limit of the above equation, one needs to be careful about the linear combination of monodromy matrices, which will give an extra $Q_{\theta}(0)$ that ensures a well defined MPS. 

In summary, we learn that the generalized dressing formula (\ref{eq:identity}) and (\ref{eq:halfintdressing})
are universal, i.e.\ they are true for any representations.
%, any inhomogeneities
%and any twists.
This implies
that, when searching for overlap formulas (even with brute force numerical
methods), it is always advisable to use the simplest possible quantum
space, since the formula is universal. For the given quantum space
the formula is modified only by the normalization pre-factors which
are easy to fit at the end.

\subsection{Universality of the generalized dressing formulas wrt.\ the twist}

Until now we focused on diagonal twists but in this section we will show that  generalized dressing formulas can be derived for 
any twist.

The monodromy matrices have $GL(4)$ symmetry
\begin{equation} \label{symmon}
T^{\lambda}(u)\left[\mathcal{R}^{\lambda}\otimes\Delta(\mathcal{R})\right]=\left[\mathcal{R}^{\lambda}\otimes\Delta(\mathcal{R})\right]T^{\lambda}(u),
\end{equation}
where $\mathcal{R}\in GL(4)$, and $\Delta(\mathcal{R})$ and $\mathcal{R}^{\lambda}$
are its representations in the quantum and auxiliary spaces, respectively.  More precisely
\begin{equation}\label{auxiliary}
\mathcal{R}=\exp\left(\sum_{i,j}x_{i,j}e_{i,j}\right),\quad\mathcal{R}^{\mathbf{6}}=\exp\left(\sum_{i,j}x_{i,j}\mathcal{E}_{i,j}\right),\quad\mathcal{R}^{\lambda}=\exp\left(\sum_{i,j}x_{i,j}E_{i,j}^{\lambda}\right),
\end{equation}
where $x_{i,j}\in\mathbb{C}$ for $i,j=1,\dots,4$ and
\begin{equation}
\Delta(\mathcal{R})=\mathcal{R}^{\mathbf{6}}\otimes\mathcal{R}^{\mathbf{6}}\otimes\dots\otimes\mathcal{R}^{\mathbf{6}}.
\label{quantum}
\end{equation}
 From the symmetry equation of the monodromy matrix,~eqn.\ (\ref{symmon})  we obtain
\begin{equation}
\mathrm{Tr}_{0}\left(T_{0}(u)G_{0}\right)\Delta(\mathcal{R})=\Delta(\mathcal{R})\mathrm{Tr}_{0}\left(T_{0}(u)\mathcal{R}_{0}^{-1}G_{0}\mathcal{R}_{0}\right),
\end{equation}
Therefore,
\begin{equation}
\mathcal{T}^{G}(u)\Delta(\mathcal{R})=\Delta(\mathcal{R})\mathcal{T}^{G_{\mathcal{R}}}(u),
\end{equation}
where $G_{\mathcal{R}}=\mathcal{R}^{-1}G\mathcal{R}$ and the superscript
denotes the twist matrix. This relation can be also generalized to
the other transfer matrices
\begin{equation}
\widehat{\mathcal{T}}^{G}(u)\Delta(\mathcal{R})=\Delta(\mathcal{R})\widehat{\mathcal{T}}^{G_{\mathcal{R}}}(u),\quad\mathcal{T}_{\lambda}^{G}(u)\Delta(\mathcal{R})=\Delta(\mathcal{R})\mathcal{T}_{\lambda}^{G_{\mathcal{R}}}(u),
\end{equation}
where
\begin{equation}
\widehat{\mathcal{T}}^{G}(u)=\mathrm{Tr}_{0}\left(\widehat{T}_{0}(u)\widehat{G}_{0}\right),\quad\mathcal{T}_{\lambda}^{G}(u)=\mathrm{Tr}_{0}\left(T_{0}^{\lambda}(u)G_{0}^{\lambda}\right)
\end{equation}
Let us introduce a rotated MPS by 
\begin{equation} \label{eq:rotMPS}
\langle \mbox{MPS}^{\mathcal{R}}|=\langle \mbox{MPS}|\Delta(\mathcal{R}),\end{equation}
If an MPS is compatible with a certain twist matrix i.e.
\begin{equation}
\langle\mathrm{MPS}|\mathcal{T}^{G}(u)=\langle\mathrm{MPS}|\widehat{\mathcal{T}}^{G}(-u),
\end{equation}
then the rotated MPS  is compatible with the rotated
twist
\begin{equation}
\langle\mathrm{MPS}^{\mathcal{R}}|\mathcal{T}^{G_{\mathcal{R}}}(u)=\langle\mathrm{MPS}^{\mathcal{R}}|\widehat{\mathcal{T}}^{G_{\mathcal{R}}}(-u).
\end{equation}
If our dressing formulas are satisfied for a pair of MPS and twist $(\langle\mathrm{MPS}|,G)$
they are also satisfied for the rotated versions $(\langle\mathrm{MPS}^{\mathcal{R}}|,G_{\mathcal{R}})$.
This means that our dressing formulas also work for non-diagonal twists.
We just need to make sure that we choose the K-matrix and MPS that
are compatible with the twist (the equations (\ref{eq:sym}) and (\ref{eq:intCond})
are satisfied). 

In practical terms, if we have a general twist and a compatible K-matrix (eq. (3.5)) then we can use a GL(4) transformation like in (4.53) to diagonalize the twist matrix G. In the rotated frame the twist matrix G is diagonal, therefore the dressing formulas are true in that frame. Finally, we can rotate back and get the dressing formulas in the original frame. These formulas will involve transfer matrices with modified twists. Any twist can be obtained from a diagonal one by a GL(4) transformation, therefore generalized dressing formulas exist for any twist.

\section{Application to the D3-D5 domain wall \label{D3D5}}

In this section we finally return to the overlap of relevance for the D3-D5 domain wall described in the introduction. Our
aim is to derive the overlap $\langle \mbox{MPS}_k | \bar{u}\rangle$ between the matrix product state given by eqns.~(\ref{intro_eqn1}) and~(\ref{intro_eqn2}) and the eigenstates of the integrable $\mathfrak{so}(6)$ spin chain. We
first exploit the $GL(4)$ symmetry of the monodromy and transfer matrices to perform a specific $GL(4)$ rotation 
which transforms the dressing formulas (\ref{eq:identity}) and  (\ref{eq:halfintdressing}) into formulas involving exactly the integrable matrix product state of interest to us. Secondly, by another $GL(4)$ rotation we relate the left and the right version of the overlap for the integer spin case. Straightforward manipulations of the dressing formulas subsequently allow us to 
recover the result previously derive by numerical investigations~\cite{DeLeeuw:2018cal}.

As in the previous section we start from the symmetry relation for the monodromy matrix, eqn. (\ref{symmon}) with a $GL(4)$ rotation
$\mathcal{R}$ represented in the auxiliary and quantum spaces as in~(\ref{auxiliary}) and~(\ref{quantum}).
In the untwisted case the transfer matrices also have $GL(4)$ symmetry:
\begin{equation}
\mathcal{T}_{\lambda}(u)\Delta(\mathcal{R})=\Delta(\mathcal{R})\mathcal{T}_{\lambda}(u).
\end{equation}
This symmetry can be applied to (\ref{eq:identity}) to get the generalized
dressing formula for the ''rotated'' MPSs
\begin{multline}
\langle \mbox{MPS}_{2s+1}^{R,\mathcal{R}}|+\langle \mbox{MPS}_{2s+1}^{L,\mathcal{R}}|=\langle\delta^{\mathcal{R}}|\mathcal{T}_{(s,s,0,0)}(-1/2)+\\
+\frac{Q_{\theta}(s)}{Q_{\theta}(1)}\langle\delta^{\mathcal{R}}|\left(\mathcal{T}_{(s-1,s-3,0,0)}(-s+1/2)-\mathcal{T}_{(s,s-2,0,0)}(-s+1/2)-\mathcal{T}_{(s-3,s-3,0,0)}(-s+1/2)\right),
\end{multline}
where the rotated MPSs are defined as in eqn.~(\ref{eq:rotMPS}) and similarly for the rotated two-site state.
The rotation for the other type of generalized dressing formula (\ref{eq:halfintdressing})
is analogous. If the original MPS and two-site state are given by the
formulas (\ref{eq:MPS}) and (\ref{eq:deltastate}) then the rotated
ones are
\begin{equation}
\begin{split}\langle \mbox{MPS}^{\mathcal{R}}| & =\sum_{i_{1},\dots,i_{2J}}\mathrm{Tr}\left[\psi_{i_{2J-1},i_{2J}}^{\mathcal{R}}(\theta_{J})\dots\psi_{i_{1},i_{2}}^{\mathcal{R}}(\theta_{1})\right]\langle i_{1},i_{2},\dots,i_{2J-1},i_{2J}|,\\
\langle\delta^{\mathcal{R}}| & =\langle\varphi^{\mathcal{R}}|^{\otimes J},
\end{split}
\end{equation}
where
\begin{equation}
\langle\varphi^{\mathcal{R}}|=\langle\varphi|\left[\mathcal{R}^{\mathbf{6}}\otimes\mathcal{R}^{\mathbf{6}}\right],\quad\psi_{i,j}^{\mathcal{R}}(u)=\sum_{k,l}\psi_{k,l}^{\mathcal{R}}(u)\left(\mathcal{R}^{\mathbf{6}}\right)_{k,i}\left(\mathcal{R}^{\mathbf{6}}\right)_{l,j}.
\end{equation}
Let us consider the $GL(4)$ transformation 
\begin{equation}
\mathcal{R}^{\mathbf{6}}=\exp\left(\frac{i\pi}{4}\left(\mathcal{E}_{1,4}+\mathcal{E}_{4,1}+\mathcal{E}_{2,3}+\mathcal{E}_{3,2}\right)\right). \label{GL4}
\end{equation}
That leads to a $\psi_{j,k}^{\mathcal{R}}(u)$ which we give in explicit form in appendix~\ref{K-matrices} and which factorizes
as follows for $u=0$
\begin{equation}
\psi_{j,k}^{\mathcal{R}}(u)=\omega_{k}^{\mathcal{R}}\omega_{j}^{\mathcal{R}},
\end{equation}
where
\begin{equation}\label{rotatedomegas}
\bar{\omega}^{\mathcal{R}}=\left\{ \omega_{Z}^{\mathcal{R}},\omega_{Y}^{\mathcal{R}},\omega_{X}^{\mathcal{R}},\omega_{\bar{X}}^{\mathcal{R}},\omega_{\bar{Y}}^{\mathcal{R}},\omega_{\bar{Z}}^{\mathcal{R}}\right\} =\left\{ S_{1}^{R}+S_{2}^{L},S_{2}^{R}+S_{1}^{L},S_{3}^{R}+S_{3}^{L},S_{3}^{R}-S_{3}^{L},S_{2}^{R}-S_{1}^{L},S_{1}^{R}-S_{2}^{L}\right\}.
\end{equation}

With the $(0,s)$ representation the rotated matrices read
\begin{equation}\label{omegaRR}
\bar{\omega}^{R,\mathcal{R}}=\left\{ \omega_{Z}^{R,\mathcal{R}},\omega_{Y}^{R,\mathcal{R}},\omega_{X}^{R,\mathcal{R}},\omega_{\bar{X}}^{R,\mathcal{R}},\omega_{\bar{Y}}^{R,\mathcal{R}},\omega_{\bar{Z}}^{R,\mathcal{R}}\right\} =\left\{ S_{1},S_{2},S_{3},S_{3},S_{2},S_{1}\right\},
\end{equation}
and  give the usual convention  
for the MPS
which describes the D3-D5 domain wall one-point functions \cite{DeLeeuw:2018cal}, cf.\ eqns.~(\ref{intro_eqn1})-(\ref{intro_eqn3}).
For the $(s,0)$ representation we  have $\langle \mbox{MPS}_{2s+1}^{L,\mathcal{R}}|$
for which
\begin{equation} \label{omegaLR}
\bar{\omega}^{L,\mathcal{R}}=\left\{ \omega_{Z}^{L,\mathcal{R}},\omega_{Y}^{L,\mathcal{R}},\omega_{X}^{L,\mathcal{R}},\omega_{\bar{X}}^{L,\mathcal{R}},\omega_{\bar{Y}}^{L,\mathcal{R}},\omega_{\bar{Z}}^{L,\mathcal{R}}\right\} =\left\{ S_{2},S_{1},S_{3},-S_{3},-S_{1},-S_{2}\right\} .
\end{equation}
We have the identity 
\begin{equation}
\sum_{a,b}\exp(\frac{i\pi}{2}S_{3})\psi_{a,b}^{L,\mathcal{R}}(u)\exp(-\frac{i\pi}{2}S_{3})\mathcal{D}_{a,j}^{\mathbf{6}}\mathcal{D}_{b,k}^{\mathbf{6}}=-\psi_{j,k}^{R,\mathcal{R}}(u),
\end{equation}
where
\begin{equation}
\mathcal{D}^{\mathbf{6}}=\exp\left(\frac{i\pi}{4}\left(\mathcal{E}_{1,1}+\mathcal{E}_{2,2}-3\mathcal{E}_{3,3}+\mathcal{E}_{4,4}\right)\right).
\end{equation}
Using this identity it is easy to show that
\begin{equation}
\langle \mbox{MPS}^{L,\mathcal{R}}|\Delta(\mathcal{D})=(-1)^{J}\langle \mbox{MPS}^{R,\mathcal{R}}|.
\end{equation}
Let us apply the this equation on a Bethe state
\begin{equation}
\langle \mbox{MPS}^{L,\mathcal{R}}|\Delta(\mathcal{D})|\bar{u}\rangle=(-1)^{J}\langle \mbox{MPS}^{R,\mathcal{R}}|\bar{u}\rangle.
\end{equation}
Since the operator $\mathcal{D}^{\mathbf{6}}$ is diagonal the co-product
$\Delta(\mathcal{D})$ acts diagonally on the Bethe states:
\begin{equation}
\Delta(\mathcal{D})|\bar{u}\rangle=\exp\left(i\pi\left(J-n_{2}+n_{3}\right)\right)=(-1)^{J+n_{2}+n_{3}},
\end{equation}
therefore the overlap of the state $\langle \mbox{MPS}^{L,\mathcal{R}}|$
can be expressed with the overlap of the other state $\langle \mbox{MPS}^{R,\mathcal{R}}|$
as
\begin{equation}
\langle \mbox{MPS}_{2s+1}^{L,\mathcal{R}}|\bar{u}\rangle=(-1)^{n_{2}+n_{3}}\langle \mbox{MPS}_{2s+1}^{R,\mathcal{R}}|\bar{u}\rangle.
\end{equation}
Assuming that the number of Bethe roots are even the MPS overlap for
integer spin can be expressed as
\begin{equation}
\langle \mbox{MPS}_{2s+1}^{R,\mathcal{R}}|\bar{u}\rangle=\langle\delta^{\mathcal{R}}|\bar{u}\rangle R_{s},
\end{equation}
where we defined the ratio as
\begin{multline}
R_{s}=\frac{1}{2}\tau_{(s,s,0,0)}(-1/2)+\\
+\frac{1}{2}\frac{Q_{\theta}(s)}{Q_{\theta}(1)}\left(\tau_{(s-1,s-3,0,0)}(-s+1/2)-\tau_{(s,s-2,0,0)}(-s+1/2)-\tau_{(s-3,s-3,0,0)}(-s+1/2)\right).
\end{multline}
Using the explicit expression for the eigenvalues (\ref{eq:eig})
and the pair structures 
\begin{equation}
Q_{\theta}(z)=Q_{\theta}(-z),\quad Q_{k}(z)=Q_{k}(-z),\quad\text{for }k=1,2,3,
\end{equation}
we can show that
\begin{equation}
R_{s}=\sum_{a=-s}^{s}Q_{\theta}(a)\frac{Q_{1}(a)}{Q_{1}(0)}\frac{Q_{2}(s+1/2)Q_{2}(1/2)}{Q_{2}(a+1/2)Q_{2}(a-1/2)}\frac{Q_{3}(a)}{Q_{3}(0)}.
\end{equation}
where the spins are integers. This result agrees with the previously
found formula \cite{DeLeeuw:2018cal}.

In the half integer case MPS overlap can be expressed as
\begin{equation}
\langle \mbox{MPS}_{2s+2}^{R,\mathcal{R}}|\bar{u}\rangle=\langle\sigma^{R,\mathcal{R}}|\bar{u}\rangle\tilde{R}_{s},
\end{equation}
where
\begin{multline}
\tilde{R}_{s}=\frac{1}{Q_{\theta}(1/2)}\tau_{(s,s,0,0)}(0)+\\
+\frac{Q_{\theta}(s+1/2)}{Q_{\theta}(1/2)^{2}}\left(\tau_{(s-1,s-2,0,0)}(-s)-\tau_{(s,s-1,0,0)}(-s)-\tau_{(s-2,s-2,0,0)}(-s)\right).
\end{multline}
Repeating the previous calculations we obtain that
\begin{equation}
\tilde{R}_{s}=\frac{1}{2}\sum_{a=-s-1/2}^{s+1/2}\frac{Q_{\theta}(a)}{Q_{\theta}(1/2)}\frac{Q_{1}(a)}{Q_{1}(1/2)}\frac{Q_{2}(s+1)Q_{2}(0)}{Q_{2}(a+1/2)Q_{2}(a-1/2)}\frac{Q_{3}(a)}{Q_{3}(1/2)},
\end{equation}
which also agrees with \cite{DeLeeuw:2018cal}.

Notice that the  detailed calculations of on-shell overlaps of this section
cannot be applied in the twisted case since the twisted transfer
matrices do not have $GL(4)$ symmetry (rather the $GL(4)$ rotations transform between different twists),
and the selection rules do no imply pairing of Bethe roots as explained in section~\ref{sec:onshell}. 
 Due to the lack of GL(4) symmetry,
it is not possible to use \textquotedblleft rotations\textquotedblright{}
to relate the overlaps $\langle\mathrm{MPS}_{2s+1}^{L}|\bar{u}\rangle$
to the overlaps $\langle\mathrm{MPS}_{2s+1}^{R}|\bar{u}\rangle$,
i.e.\ the dressing formula only gives the sum of the overlaps $\langle\mathrm{MPS}_{2s+1}^{L}|\bar{u}\rangle+\langle\mathrm{MPS}_{2s+1}^{R}|\bar{u}\rangle$. Furthermore, 
due to the lack of pair structure, the linear combination $R_{s}$ does
not simplify.

\section{Discussion and Conclusion\label{conclusion}}

We have simplified, sharpened and extended our twisted Yangian approach to the computation of overlaps between integrable matrix product states and Bethe eigenstates. In particular, we have incorporated inhomogeneities and twists in our formalism. We have concentrated on particular matrix product states of the integrable  ${SO}(6)$ spin chain which play a central role for the computation of one-point functions in the D3-D5 domain wall version of ${\cal N}=4$ SYM. These matrix product states originate from evaluation representations of the twisted Yangian $Y^+(4)$. The generic evaluation representations of $Y^+(4)$ are characterized by two spin quantum numbers, $s_R$ and $s_L$, of a right and 
a left version of $\mathfrak{sl}_2$ but only those for which one of the $\mathfrak{sl}_2$ representations is trivial, denoted as  $V_R(s)$ and $V_L(s)$,
appear in the D3-D5 domain wall problem.  Our main result consists of two dressing formulas, one for integer and one for half-integer $s$, which express the sum of the two matrix product states corresponding to respectively $V_R(s)$ and $V_L(s)$  as a linear combination of fused transfer matrices acting on the MPS of the trivial representation.
 We stress that these dressing formulas are highly universal. They are valid for any choice of twists and inhomogeneities and they also hold for any choice of quantum space for the spin chain, not just the space corresponding to the fundamental representation which is the one which appears in ${\cal N}=4$ SYM. By means of
these dressing formulas we managed to fill the final gap in the analytical understanding of the overlaps in the $\mathfrak{so}(6)$
sector of ${\cal N}=4$ SYM. 

Provided one can determine the overlap between the Bethe eigenstates and the MPS corresponding to the trivial representation of $Y^+(4)$,  and provided that the left and the right version of the general MPS can be related, the dressing formulas give access to a closed overlap formula for any MPS of evaluation representation type. In the case where the twist vanishes 
the former overlap can be read off from~\cite{Gombor:2021hmj,Gombor:2023bez} where the overlap was determined for
any two site product state for any GL(N) spin chain.
Devising a method which allows one to find the same
overlap in the presence of twists constitutes an interesting open problem which could possibly be approached by means of
the functional SoV approach~\cite{Sklyanin:1995bm,Gromov:2016itr,Ekhammar:2023iph}. In the case where twists are 
present, however, there is no simple relation between the left and the right version of the MPS. Yet our dressing formulas give rise to a non-trivial summation rule for right and left MPS overlaps. Furthermore the overlaps with the trivial representation 
in the twisted case may simplify if one introduces the full $\mathcal{Q}$-system~\cite{Tsuboi:1998ne,Gromov:2019icz,Kazakov:2018ugh,Levkovich-Maslyuk:2019awk} and not only the limited set of $Q$-functions used here.

One may wonder if the general evaluation representations $V(s_R,s_L)$ and their corresponding integrable matrix product states have an interpretation within the AdS/CFT correspondence. An obvious place to search for such an interpretation would be the non-supersymmetric D3-D7 domain wall version of ${\cal N}=4$ SYM, which is 
characterized by the scalar fields attaining vevs which are given by two different representations of $\mathfrak{su}(2)$~\cite{Kristjansen:2012tn}. A glance at the eqn.~(\ref{rotatedomegas}) immediately shows that this interpretation can not be correct as the mapping of the spin chain states to a set of three complex scalar fields
and their complex conjugates (or rather six real scalar fields)  only works for the representations $V_R(s)$ and $V_L(s)$. This is in accordance with the fact .tethat the former
D3-D7 domain wall version of ${\cal N}=4$ SYM was shown to be non integrable~\cite{deLeeuw:2019sew}.

It is noteworthy that one can encounter a situation where the homogeneous limit of the dressing formulas is subtle, cf.\ eqn.\ (\ref{subtle}). Understanding under which circumstances
this happens is an interesting open problem.

 It would likewise be interesting to extend the results of the present paper to other spin chains and in particular other reflection algebras. The extension of overlap formulas to other
 reflection algebras, albeit in the case of vanishing twists, is addressed in the paper~\cite{Gombor:2024iix}. Whereas each reflection algebra 
 has to be treated independently  the resulting formulas are highly universal.

\acknowledgments

T.G. was supported by the NKFIH grant PD142929 and the János Bolyai Research Scholarship of the Hungarian Academy of Science.
C.K.\ was supported in part by DFF-FNU through grant number 1026-00103B. V.M.\ was supported by the STFC under the grant ST/X508809/1. X.Q.\ was supported by CSC through grant number 202207940019, he also thanks Yau center of Southeast University for hospitality. 

\appendix

\section{K-matrices and  boundary states for the evaluation representations \label{K-matrices}}
In this section we collect the $K$-matrices for the evaluation
representations discussed in section~\ref{evaluation}. The $K$-matrix for the 
general representation $V(s_L, s_R)$ reads
\begin{equation}
\mathbf{K}(u)=\left(\begin{array}{cccc}
1+\frac{2}{2u+1}(S_{3}^{R}+S_{3}^{L}) & \frac{2\sqrt{2}}{2u+1}S_{+}^{L} & \frac{2\sqrt{2}}{2u+1}S_{+}^{R} & 0\\
\frac{2\sqrt{2}}{2u+1}S_{-}^{L} & 1+\frac{2}{2u+1}(S_{3}^{R}-S_{3}^{L}) & 0 & -\frac{2\sqrt{2}}{2u+1}S_{+}^{R}\\
\frac{2\sqrt{2}}{2u+1}S_{-}^{R} & 0 & 1-\frac{2}{2u+1}(S_{3}^{R}-S_{3}^{L}) & -\frac{2\sqrt{2}}{2u+1}S_{+}^{L}\\
0 & -\frac{2\sqrt{2}}{2u+1}S_{-}^{R} & -\frac{2\sqrt{2}}{2u+1}S_{-}^{L} & 1-\frac{2}{2u+1}(S_{3}^{R}+S_{3}^{L})
\end{array}\right),
\end{equation}
and the corresponding fused two-site state operators, $\psi_{a,b}(u)$ take the form
\begin{multline}
\psi_{a,b}(u)=\omega_{b}\omega_{a}-u^{2}\varphi_{a,b}+\\
u\left(\begin{array}{cccccc}
0 & 0 & \sqrt{2}S_{-}^{R} & \sqrt{2}S_{-}^{R} & 0 & -2S_{3}^{R}+C(u)\\
0 & 0 & \sqrt{2}S_{-}^{L} & -\sqrt{2}S_{-}^{L} & 2S_{3}^{L}+C(u) & 0\\
-\sqrt{2}S_{-}^{R} & -\sqrt{2}S_{-}^{L} & 0 & C(u) & -\sqrt{2}S_{+}^{L} & \sqrt{2}S_{+}^{R}\\
-\sqrt{2}S_{-}^{R} & \sqrt{2}S_{-}^{L} & C(u) & 0 & \sqrt{2}S_{+}^{L} & \sqrt{2}S_{+}^{R}\\ \label{fusedtwosite}
0 & -2S_{3}^{L}+C(u) & \sqrt{2}S_{+}^{L} & -\sqrt{2}S_{+}^{L} & 0 & 0\\
2S_{3}^{R}+C(u) & 0 & -\sqrt{2}S_{+}^{R} & -\sqrt{2}S_{+}^{R} & 0 & 0
\end{array}\right)_{a,b},
\end{multline}
where $a,b=1,\dots,6$ and
\begin{align}
C(u) & =\frac{1}{u+1}\left(\left(S^{L}\right)^{2}-\left(S^{R}\right)^{2}\right),\quad\left(S^{L/R}\right)^{2}=\sum_{k=1}^{3}\left(S_{k}^{L/R}\right)^{2},\\
\bar{\omega} & =\left\{ \omega_{Z},\omega_{Y},\omega_{X},\omega_{\bar{X}},\omega_{\bar{Y}},\omega_{\bar{Z}}\right\} =\left\{ \sqrt{2}S_{-}^{R},\sqrt{2}S_{-}^{L},S_{3}^{R}+S_{3}^{L},S_{3}^{R}-S_{3}^{L},-\sqrt{2}S_{+}^{L},\sqrt{2}S_{+}^{R}\right\}.
\end{align}
Finally,
\begin{equation} \label{phiij}
\varphi_{a,b}=\left(\begin{array}{cccccc}
0 & 0 & 0 & 0 & 0 & 1\\
0 & 0 & 0 & 0 & -1 & 0\\
0 & 0 & 1 & 0 & 0 & 0\\
0 & 0 & 0 & 1 & 0 & 0\\
0 & -1 & 0 & 0 & 0 & 0\\
1 & 0 & 0 & 0 & 0 & 0
\end{array}\right)_{a,b}.
\end{equation}
Applying the GL(4) rotation given in eqn.~(\ref{GL4}) the fused two site operator~(\ref{fusedtwosite}) can be brought on the form
\begin{multline}
\psi_{j,k}^{\mathcal{R}}(u)=\omega_{k}^{\mathcal{R}}\omega_{j}^{\mathcal{R}}-u^{2}\varphi_{j,k}^{\mathcal{R}}+\\
u\left(\begin{array}{cccccc}
0 & -i\left(S_{3}^{L}-S_{3}^{R}\right) & i\left(S_{1}^{L}-S_{2}^{R}\right) & -i\left(S_{1}^{L}+S_{2}^{R}\right) & i\left(S_{3}^{L}+S_{3}^{R}\right) & C(u)\\
i\left(S_{3}^{L}-S_{3}^{R}\right) & 0 & -i\left(S_{2}^{L}-S_{1}^{R}\right) & i\left(S_{2}^{L}+S_{1}^{R}\right) & C(u) & -i\left(S_{3}^{L}+S_{3}^{R}\right)\\
-i\left(S_{1}^{L}-S_{2}^{R}\right) & i\left(S_{2}^{L}-S_{1}^{R}\right) & 0 & C(u) & -i\left(S_{2}^{L}+S_{1}^{R}\right) & i\left(S_{1}^{L}+S_{2}^{R}\right)\\
i\left(S_{1}^{L}+S_{2}^{R}\right) & -i\left(S_{2}^{L}+S_{1}^{R}\right) & C(u) & 0 & i\left(S_{2}^{L}-S_{1}^{R}\right) & -i\left(S_{1}^{L}-S_{2}^{R}\right)\\
-i\left(S_{3}^{L}+S_{3}^{R}\right) & C(u) & i\left(S_{2}^{L}+S_{1}^{R}\right) & -i\left(S_{2}^{L}-S_{1}^{R}\right) & 0 & i\left(S_{3}^{L}-S_{3}^{R}\right)\\
C(u) & i\left(S_{3}^{L}+S_{3}^{R}\right) & -i\left(S_{1}^{L}+S_{2}^{R}\right) & i\left(S_{1}^{L}-S_{2}^{R}\right) & -i\left(S_{3}^{L}-S_{3}^{R}\right) & 0
\end{array}\right)_{j,k},
\end{multline}
where the $\bar{\omega}^{\mathcal{R}}$ was given in eqn.\ (\ref{rotatedomegas})
and
\begin{equation}
\varphi_{a,b}^{\mathcal{R}}=\left(\begin{array}{cccccc}
1 & 0 & 0 & 0 & 0 & 0\\
0 & 1 & 0 & 0 & 0 & 0\\
0 & 0 & 1 & 0 & 0 & 0\\
0 & 0 & 0 & 1 & 0 & 0\\
0 & 0 & 0 & 0 & 1 & 0\\
0 & 0 & 0 & 0 & 0 & 1
\end{array}\right)_{a,b}.
\end{equation}

Specializing to the representation $V(0,s)\equiv V_R(s)$ we get the $K$-matrix
\begin{equation}
\mathbf{K}^{R}(u)=\left(\begin{array}{cccc}
1+\frac{2}{2u+1}S_{3} & 0 & \frac{2\sqrt{2}}{2u+1}S_{+} & 0\\
0 & 1+\frac{2}{2u+1}S_{3} & 0 & -\frac{2\sqrt{2}}{2u+1}S_{+}\\
\frac{2\sqrt{2}}{2u+1}S_{-} & 0 & 1-\frac{2}{2u+1}S_{3} & 0\\
0 & -\frac{2\sqrt{2}}{2u+1}S_{-} & 0 & 1-\frac{2}{2u+1}S_{3}
\end{array}\right),
\end{equation}
and similarly for $V(s,0)\equiv V_L(s)$
\begin{equation}
\mathbf{K}^{L}(u)=\left(\begin{array}{cccc}
1+\frac{2}{2u+1}S_{3} & \frac{2\sqrt{2}}{2u+1}S_{+} & 0 & 0\\
\frac{2\sqrt{2}}{2u+1}S_{-} & 1-\frac{2}{2u+1}S_{3} & 0 & 0\\
0 & 0 & 1+\frac{2}{2u+1}S_{3} & -\frac{2\sqrt{2}}{2u+1}S_{+}\\
0 & 0 & -\frac{2\sqrt{2}}{2u+1}S_{-} & 1-\frac{2}{2u+1}S_{3}
\end{array}\right).
\end{equation}
The corresponding  $\bar{\omega}^{R,\mathcal{R}}$ and $\bar{\omega}^{L,\mathcal{R}}$ were given in eqn.~(\ref{omegaRR}) and~(\ref{omegaLR}).

\section{Derivation of dressing formulas for integer spin \label{integerbranching}}
In this section we explicitly derive the dressing formulas for $s=1$, $s=2$ and $s=3$ by going through the iterative
procedure described in section~\ref{recursion}.  The key idea is to express $L(s,s,0,0|1/2)$ in terms of $V_R(s)\oplus V_L(s)$.

\subsection{Formulas for $s=1$}

Let us start with $s=1$. For the $\mathfrak{gl}_{4}$ representation
$\lambda=(1,1,0,0)$ we have six states and the $\mathfrak{gl}_{4}$
and $\mathfrak{o}_{4}$ weights are shown in table \ref{tab:1100}.
The state $(0,0,1,1)$ is highest weight state and the highest weight
is the same as $V_{R}(1)$. States of the representations $V_{R}(1)$
and $V_{L}(1)$ in the $\mathfrak{\ensuremath{o}_{4}}$ weight space
are shown in table \ref{tab:VR1}. From the patterns we are lead to the
following conjecture
\begin{equation}
L((1,1,0,0)|1/2)=V_{R}(1)\oplus V_{L}(1).\label{eq:LVV}
\end{equation}
It can easily be shown that this is true. The state $(0,1,0,1)$ is
also a highest weight state and the highest weights are
\begin{equation}
\mu_{4}(u)=\frac{u-1/2}{u+1/2},\quad\mu_{3}(u)=\frac{u+3/2}{u+1/2}.
\end{equation}
Therefore,
\begin{equation}
P_{1}(u)=\frac{u-1/2}{u+3/2},\quad P_{2}(u)=\frac{(u+1/2)(u-3/2)}{(u-1/2)(u+3/2)}.
\end{equation}
This agrees with the ratio of highest weights of $V_{L}(1)$, therefore
we just showed (\ref{eq:LVV}). It means that we obtained the generalized
dressing formula for $s=1$:
\begin{equation}
\langle \mbox{MPS}_{1}^{R}|+\langle \mbox{MPS}_{1}^{L}|=\langle\delta|\mathcal{T}_{(1,1,0,0)}(-1/2).
\end{equation}

\begin{table}
\begin{centering}
\begin{tabular}{|c|c|c|c|}
\hline 
$s_{R}\backslash s_{L}$ & 1 & 0 & -1\tabularnewline
\hline 
\hline 
1 &  & (0,0,1,1) & \tabularnewline
\hline 
0 & (0,1,0,1) & (1,0,0,1),(0,1,1,0) & (1,0,1,0)\tabularnewline
\hline 
-1 &  & (1,1,0,0) & \tabularnewline
\hline 
\end{tabular}$\to$%
\begin{tabular}{|c|c|c|c|}
\hline 
$s_{R}\backslash s_{L}$ & 1 & 0 & -1\tabularnewline
\hline 
\hline 
1 &  & $\times1$ & \tabularnewline
\hline 
0 & $\times1$ & $\times2$ & $\times1$\tabularnewline
\hline 
-1 &  & $\times1$ & \tabularnewline
\hline 
\end{tabular}
\par\end{centering}
\caption{The $\mathfrak{gl}_{4}$ and $\mathfrak{o}_{4}$ weights of the representation
$L(1,1,0,0)$ }

\label{tab:1100}
\end{table}

\begin{table}
\begin{centering}
\begin{tabular}{|c|c|c|c|}
\hline 
$s_{R}\backslash s_{L}$ & 1 & 0 & -1\tabularnewline
\hline 
\hline 
1 &  &  & \tabularnewline
\hline 
0 & $\times1$ & $\times1$ & $\times1$\tabularnewline
\hline 
-1 &  &  & \tabularnewline
\hline 
\end{tabular}\qquad{}%
\begin{tabular}{|c|c|c|c|}
\hline 
$s_{R}\backslash s_{L}$ & 1 & 0 & -1\tabularnewline
\hline 
\hline 
1 &  & $\times1$ & \tabularnewline
\hline 
0 &  & $\times1$ & \tabularnewline
\hline 
-1 &  & $\times1$ & \tabularnewline
\hline 
\end{tabular}
\par\end{centering}
\caption{States of the representations $V_{R}(1)$ and $V_{L}(1)$ in the $\mathfrak{\ensuremath{o}_{4}}$
weight space.}

\label{tab:VR1}
\end{table}

\subsection{Formulas for $s=2$}

Let us continue with $s=2$. For the $\mathfrak{gl}_{4}$ representation
$\lambda=(2,2,0,0)$ we have 20 states and the $\mathfrak{gl}_{4}$
and $\mathfrak{o}_{4}$ weights are shown in table \ref{tab:2200}.
The state $(0,0,2,2)$ is a highest weight state and the highest weight
is the same as for $V_{R}(2)$. The state $(0,2,0,2)$ is also a highest
weight state and the highest weights are
\begin{equation}
\begin{split}\mu_{4}(u) & =\frac{u-3/2}{u+1/2},\quad\mu_{3}(u)=\frac{u+5/2}{u+1/2},\\
P_{1}(u) & =\frac{u-3/2}{u+5/2},\quad P_{2}(u)=\frac{(u+1/2)(u-5/2)}{(u-1/2)(u+5/2)}.
\end{split}
\end{equation}
This agrees with the ratio of highest weights of $V_{L}(2).$ States
of the representations $V_{R}(2)$ and $V_{L}(2)$ in the $\mathfrak{\ensuremath{o}_{4}}$
weight space are shown in table \ref{tab:VR2}. From the pattern
\begin{equation}
\begin{array}{ccccc}
 &  & 1\\
 & 1 & 2 & 1\\
1 & 2 & 4 & 2 & 1\\
 & 1 & 2 & 1\\
 &  & 1
\end{array}=\begin{array}{ccccc}
 &  & 1\\
 &  & 1\\
 &  & 1\\
 &  & 1\\
 &  & 1
\end{array}\oplus\begin{array}{ccccc}
\\
\\
1 & 1 & 1 & 1 & 1\\
\\
\\
\end{array}\oplus\begin{array}{ccccc}
\\
 & 1 & 1 & 1\\
 & 1 & 2 & 1\\
 & 1 & 1 & 1\\
\\
\end{array}
\end{equation}
 we are lead to the following conjecture
\begin{equation}
L((2,2,0,0)|1/2)=V_{R}(2)\oplus V_{L}(2)\oplus L((2,0,0,0)|\xi).\label{eq:LVV-1}
\end{equation}
It can easily be shown that this is true. The dimension of the factor
space  \\
\mbox{$V^{(2)}=L((2,2,0,0)|1/2)\backslash\left[V_{R}(2)\oplus V_{L}(2)\right]$} is 10~which agrees with the dimension of $L((2,0,0,0)|\xi)$. The
state $(0,1,1,2)$ is a highest weight state in the factor space $V^{(2)}$
and its highest weights are
\[
\mu_{4}(u)=\frac{u-3/2}{u+1/2},\quad\mu_{3}(u)=\frac{(u-3/2)(u+3/2)}{(u-1/2)(u+1/2)},
\]
therefore
\begin{equation}
P_{1}(u)=\frac{u-1/2}{u+3/2},\quad P_{2}(u)=1.\label{eq:PP}
\end{equation}
From (\ref{eq:dressedP}) we can see that (\ref{eq:PP}) agrees with the
ratio of highest weights of the representation $L((2,0,0,0)|3/2)$,
therefore we just showed that
\begin{equation}
L((2,2,0,0)|1/2)=V_{R}(2)\oplus V_{L}(2)\oplus L((2,0,0,0)|3/2).
\end{equation}
 This means that we obtained the generalized dressing formula for $s=2$:
\begin{equation}
\langle \mbox{MPS}_{2}^{R}|+\langle \mbox{MPS}_{2}^{L}|=\langle\delta|\mathcal{T}_{(2,2,0,0)}(-1/2)-\frac{Q_{\theta}(2)}{Q_{\theta}(1)}\langle\delta|\mathcal{T}_{(2,0,0,0)}(-3/2),
\end{equation}
where the normalization factor was fitted numerically.

\begin{table}
\begin{centering}
\begin{tabular}{|c|c|c|c|c|c|}
\hline 
$s_{R}\backslash s_{L}$ & 2 & 1 & 0 & -1 & -2\tabularnewline
\hline 
2 &  &  & (0,0,2,2) &  & \tabularnewline
\hline 
1 &  & (0,1,1,2) & (1,0,1,2),(0,1,2,1) & (1,0,2,1) & \tabularnewline
\hline 
0 & (0,2,0,2) & (0,2,1,1),(1,1,0,2) & 2$\times$(1,1,1,1),(0,2,2,0),(2,0,0,2) & (1,1,2,0),(2,0,1,1) & (2,0,2,0)\tabularnewline
\hline 
-1 &  & (1,2,0,1) & (2,1,0,1),(1,2,1,0) & (2,1,1,0) & \tabularnewline
\hline 
-2 &  &  & (2,2,0,0) &  & \tabularnewline
\hline 
\end{tabular}$\to$
\par\end{centering}
\begin{centering}
\begin{tabular}{|c|c|c|c|c|c|}
\hline 
$s_{R}\backslash s_{L}$ & 2 & 1 & 0 & -1 & -2\tabularnewline
\hline 
\hline 
2 &  &  & $\times1$ &  & \tabularnewline
\hline 
1 &  & $\times1$ & $\times2$ & $\times1$ & \tabularnewline
\hline 
0 & $\times1$ & $\times2$ & $\times4$ & $\times2$ & $\times1$\tabularnewline
\hline 
-1 &  & $\times1$ & $\times2$ & $\times1$ & \tabularnewline
\hline 
-2 &  &  & $\times1$ &  & \tabularnewline
\hline 
\end{tabular}
\par\end{centering}
\caption{The $\mathfrak{gl}_{4}$ and $\mathfrak{o}_{4}$ weights of the representation
$L(2,2,0,0)$}

\label{tab:2200}
\end{table}
\begin{table}
\begin{centering}
\begin{tabular}{|c|c|c|c|c|c|}
\hline 
$s_{R}\backslash s_{L}$ & 2 & 1 & 0 & -1 & -2\tabularnewline
\hline 
2 &  &  &  &  & \tabularnewline
\hline 
1 &  & (0,0,0,2) & (0,0,1,1) & (0,0,2,0) & \tabularnewline
\hline 
0 &  & (0,1,0,1) & (1,0,0,1),(0,1,1,0) & (1,0,1,0) & \tabularnewline
\hline 
-1 &  & (0,2,0,0) & (1,1,0,0) & (2,0,0,0) & \tabularnewline
\hline 
-2 &  &  &  &  & \tabularnewline
\hline 
\end{tabular}$\to$%
\begin{tabular}{|c|c|c|c|c|c|}
\hline 
$s_{R}\backslash s_{L}$ & 2 & 1 & 0 & -1 & -2\tabularnewline
\hline 
\hline 
2 &  &  &  &  & \tabularnewline
\hline 
1 &  & $\times1$ & $\times1$ & $\times1$ & \tabularnewline
\hline 
0 &  & $\times1$ & $\times2$ & $\times1$ & \tabularnewline
\hline 
-1 &  & $\times1$ & $\times1$ & $\times1$ & \tabularnewline
\hline 
-2 &  &  &  &  & \tabularnewline
\hline 
\end{tabular}
\par\end{centering}
\caption{The $\mathfrak{gl}_{4}$ and $\mathfrak{o}_{4}$ weights of the representation
$L(2,0,0,0)$}

\label{tab:2000}
\end{table}

\begin{table}
\begin{centering}
\begin{tabular}{|c|c|c|c|c|c|}
\hline 
$s_{R}\backslash s_{L}$ & 2 & 1 & 0 & -1 & -2\tabularnewline
\hline 
\hline 
2 &  &  &  &  & \tabularnewline
\hline 
1 &  &  &  &  & \tabularnewline
\hline 
0 & $\times1$ & $\times1$ & $\times1$ & $\times1$ & $\times1$\tabularnewline
\hline 
-1 &  &  &  &  & \tabularnewline
\hline 
-2 &  &  &  &  & \tabularnewline
\hline 
\end{tabular}\qquad{}%
\begin{tabular}{|c|c|c|c|c|c|}
\hline 
$s_{R}\backslash s_{L}$ & 2 & 1 & 0 & -1 & -2\tabularnewline
\hline 
\hline 
2 &  &  & $\times1$ &  & \tabularnewline
\hline 
1 &  &  & $\times1$ &  & \tabularnewline
\hline 
0 &  &  & $\times1$ &  & \tabularnewline
\hline 
-1 &  &  & $\times1$ &  & \tabularnewline
\hline 
-2 &  &  & $\times1$ &  & \tabularnewline
\hline 
\end{tabular}
\par\end{centering}
\caption{States of the representations $V_{R}(2)$ and $V_{L}(2)$ in the $\mathfrak{\ensuremath{o}_{4}}$
weight space.}

\label{tab:VR2}
\end{table}

\subsection{Formulas for $s=3$}

Let us continue with $s=3$. For the $\mathfrak{gl}_{4}$ representation
$\lambda=(3,3,0,0)$ we have 50 states and the $\mathfrak{gl}_{4}$
and $\mathfrak{o}_{4}$ weights are shown in the table \ref{tab:3300}.
The states $(0,0,3,3)$ and $(0,3,0,3)$ are highest weight states
which correspond to invariant subspaces $V_{R}(3)$ and $V_{L}(3)$.
Let us define the factor space $V^{(2)}=L((3,3,0,0)|1/2)\backslash\left[V_{R}(3)\oplus V_{L}(3)\right]$,
i.e.
\begin{equation}
L((3,3,0,0)|1/2)=V_{R}(3)\oplus V_{L}(3)\oplus V^{(2)}.
\end{equation}
 The states of $V^{(2)}$ in the $\mathfrak{o}_{4}$ weight space
look like
\begin{equation}
\begin{array}{ccccccc}
 &  &  & 1\\
 &  & 1 & 2 & 1\\
 & 1 & 2 & 4 & 2 & 1\\
1 & 2 & 4 & 6 & 4 & 2 & 1\\
 & 1 & 2 & 4 & 2 & 1\\
 &  & 1 & 2 & 1\\
 &  &  & 1
\end{array}-\left(\begin{array}{ccccccc}
 &  &  & 1\\
 &  &  & 1\\
 &  &  & 1\\
 &  &  & 1\\
 &  &  & 1\\
 &  &  & 1\\
 &  &  & 1
\end{array}\oplus\begin{array}{ccccccc}
\\
\\
\\
1 & 1 & 1 & 1 & 1 & 1 & 1\\
\\
\\
\\
\end{array}\right)=\begin{array}{ccccccc}
\\
 &  & 1 & 1 & 1\\
 & 1 & 2 & 3 & 2 & 1\\
 & 1 & 3 & 4 & 3 & 1\\
 & 1 & 2 & 3 & 2 & 1\\
 &  & 1 & 1 & 1\\
\\
\end{array}.
\end{equation}
 The state (0,1,2,3) is a highest weight state in the factor space $V^{(2)}$
and the corresponding highest weights are
\begin{equation}
\begin{split}\mu_{4}^{(2)}(u)= & \frac{u-5/2}{u+1/2},\quad\mu_{3}^{(2)}(u)=\frac{(u-5/2)(u+3/2)}{(u-1/2)(u+1/2)},\\
P_{1}^{(2)}(u)= & \frac{u-1/2}{u+3/2},\quad P_{2}^{(2)}(u)=\frac{(u+5/2)(u-3/2)}{(u-5/2)(u+3/2)},
\end{split}
\label{eq:PP2}
\end{equation}
which agree with the highest weights of $L((3,1,0,0)|5/2)$ (see
(\ref{eq:dressedP})). Since the dimension of $L((3,1,0,0)|5/2)$
is bigger then $V^{(2)}$ we have the decomposition
\begin{equation}
L((3,1,0,0)|5/2)=V^{(2)}\oplus V^{(3)}.
\end{equation}
The states of $L((3,1,0,0)|5/2)$ are shown in table \ref{tab:3100} and
the states of $V^{(3)}$ in the $\mathfrak{o}_{4}$ weights
space therefore look like
\begin{equation}
\begin{array}{ccccccc}
\\
 &  & 1 & 1 & 1\\
 & 1 & 3 & 4 & 3 & 1\\
 & 1 & 4 & 5 & 4 & 1\\
 & 1 & 3 & 4 & 3 & 1\\
 &  & 1 & 1 & 1\\
\\
\end{array}-\begin{array}{ccccccc}
\\
 &  & 1 & 1 & 1\\
 & 1 & 2 & 3 & 2 & 1\\
 & 1 & 3 & 4 & 3 & 1\\
 & 1 & 2 & 3 & 2 & 1\\
 &  & 1 & 1 & 1\\
\\
\end{array}=\begin{array}{ccccccc}
\\
\\
 &  & 1 & 1 & 1\\
 &  & 1 & 1 & 1\\
 &  & 1 & 1 & 1\\
\\
\\
\end{array}.
\end{equation}
There are three states $|\Lambda_{1}\rangle$, $|\Lambda_{2}\rangle$,
$|\Lambda_{3}\rangle$ in $L((3,1,0,0)|5/2)$ which have $(1,1)$
spins and the corresponding GT-patterns are
\begin{equation}
\Lambda_{1}=\begin{array}{ccccccc}
3 &  & 1 &  & 0 &  & 0\\
 & 2 &  & 0 &  & 0\\
 &  & 1 &  & 0\\
 &  &  & 0
\end{array},\quad\Lambda_{2}=\begin{array}{ccccccc}
3 &  & 1 &  & 0 &  & 0\\
 & 1 &  & 1 &  & 0\\
 &  & 1 &  & 0\\
 &  &  & 0
\end{array},\quad\Lambda_{3}=\begin{array}{ccccccc}
3 &  & 1 &  & 0 &  & 0\\
 & 1 &  & 0 &  & 0\\
 &  & 1 &  & 0\\
 &  &  & 1
\end{array}.
\end{equation}
The state $|\Lambda_{1}\rangle+\frac{1}{4}|\Lambda_{3}\rangle$ is
a highest weight state which generates the invariant subspace $V^{(3)}$.
The corresponding highest weights are
\begin{equation}
\begin{split}\mu_{4}^{(3)}(u)= & \frac{u+1/2}{u+5/2},\quad\mu_{3}^{(3)}(u)=1,\\
P_{1}^{(3)}(u)= & \frac{u+1/2}{u+5/2},\quad P_{2}^{(3)}(u)=1,
\end{split}
\label{eq:PP2-1}
\end{equation}
which agree with the highest weights of $L((2,0,0,0)|5/2)$. Since
the dimension of $L((2,0,0,0)|5/2)$ is 10 but the subspace $V^{(3)}$
has dimension 9, we have the following decomposition
\begin{equation}
L((2,0,0,0)|5/2)=V^{(3)}\oplus V^{(4)}.
\end{equation}
The states of $V^{(4)}$ in the $\mathfrak{o}_{4}$ weights space
look like
\begin{equation}
\begin{array}{ccccccc}
\\
\\
 &  & 1 & 1 & 1\\
 &  & 1 & 2 & 1\\
 &  & 1 & 1 & 1\\
\\
\\
\end{array}-\begin{array}{ccccccc}
\\
\\
 &  & 1 & 1 & 1\\
 &  & 1 & 1 & 1\\
 &  & 1 & 1 & 1\\
\\
\\
\end{array}=\begin{array}{ccccccc}
\\
\\
\\
 &  &  & 1,\\
\\
\\
\\
\end{array}
\end{equation}
therefore $V^{(4)}$ is the singlet representation. Summarizing 
the results, the generalized dressing formula for $s=3$ can be written
as
\begin{multline}
\langle \mbox{MPS}_{3}^{R}|+\langle \mbox{MPS}_{3}^{L}|=\\
\langle\delta|\mathcal{T}_{(3,3,0,0)}(-1/2)+\frac{Q_{\theta}(3)}{Q_{\theta}(1)}\langle\delta|\left(\mathcal{T}_{(2,0,0,0)}(-5/2)-\mathcal{T}_{(3,1,0,0)}(-5/2)-\mathcal{T}_{(0,0,0,0)}(-5/2)\right),
\end{multline}
where the normalization factor was fitted numerically.

\begin{table}
\begin{centering}
\begin{tabular}{|c|c|c|c|c|c|c|c|}
\hline 
$s_{R}\backslash s_{L}$ & 3 & 2 & 1 & 0 & -1 & -2 & -3\tabularnewline
\hline 
\hline 
3 &  &  &  & $\times1$ &  &  & \tabularnewline
\hline 
2 &  &  & $\times1$ & $\times2$ & $\times1$ &  & \tabularnewline
\hline 
1 &  & $\times1$ & $\times2$ & $\times4$ & $\times2$ & $\times1$ & \tabularnewline
\hline 
0 & $\times1$ & $\times2$ & $\times4$ & $\times6$ & $\times4$ & $\times2$ & $\times1$\tabularnewline
\hline 
-1 &  & $\times1$ & $\times2$ & $\times4$ & $\times2$ & $\times1$ & \tabularnewline
\hline 
-2 &  &  & $\times1$ & $\times2$ & $\times1$ &  & \tabularnewline
\hline 
-3 &  &  &  & $\times1$ &  &  & \tabularnewline
\hline 
\end{tabular}
\par\end{centering}
\caption{The $\mathfrak{gl}_{4}$ and $\mathfrak{o}_{4}$ weights of the representation
$L(3,3,0,0)$}

\label{tab:3300}
\end{table}

\begin{table}
\begin{centering}
\begin{tabular}{|c|c|c|c|c|c|c|c|}
\hline 
$s_{R}\backslash s_{L}$ & 3 & 2 & 1 & 0 & -1 & -2 & -3\tabularnewline
\hline 
\hline 
3 &  &  &  &  &  &  & \tabularnewline
\hline 
2 &  &  & $\times1$ & $\times1$ & $\times1$ &  & \tabularnewline
\hline 
1 &  & $\times1$ & $\times3$ & $\times4$ & $\times3$ & $\times1$ & \tabularnewline
\hline 
0 &  & $\times1$ & $\times4$ & $\times5$ & $\times4$ & $\times1$ & \tabularnewline
\hline 
-1 &  & $\times1$ & $\times3$ & $\times4$ & $\times3$ & $\times1$ & \tabularnewline
\hline 
-2 &  &  & $\times1$ & $\times1$ & $\times1$ &  & \tabularnewline
\hline 
-3 &  &  &  &  &  &  & \tabularnewline
\hline 
\end{tabular}
\par\end{centering}
\caption{The $\mathfrak{gl}_{4}$ and $\mathfrak{o}_{4}$ weights of the representation
$L(3,1,0,0)$}

\label{tab:3100}
\end{table}

\subsection{Formulas for general integer $s$}

Based on the previous analysis we have the following conjecture for the representation
embeddings
\begin{equation}
\begin{split}L((s,s,0,0)|1/2) & =V_{R}(s)\oplus V_{L}(s)\oplus V^{(2)},\\
L((s,s-2,0,0)|s-1/2) & =V^{(2)}\oplus V^{(3)},\\
L((s-1,s-3,0,0)|s-1/2) & =V^{(3)}\oplus L((s-3,s-3,0,0)|s-1/2).
\end{split}
\label{eq:branching}
\end{equation}
The dimensions of the representations are
\begin{equation}
\begin{split}V_{R}(s)\oplus V_{L}(s) & \to2(2s+1),\\
L((s,s,0,0)|1/2) & \to\frac{(s+1)(s+2)^{2}(s+3)}{12},\\
L((s,s-2,0,0)|1/2) & \to\frac{(s-1)s(s+2)(s+3)}{4}.
\end{split}
\end{equation}
From the first two equations we can express the dimensions of the $V^{(2)}$ and
$V^{(3)}$ spaces as
\begin{equation}
\begin{split}V^{(2)} & \to\frac{(s-1)(s^{3}+9s^{2}+32s+12)}{12},\\
V^{(3)} & \to\frac{(s-2)(s-1)(5s^{2}+5s+3)}{6},
\end{split}
\end{equation}
and the dimensions in the third equation are consistent with this. This is a strong
indication that the formulas are correct. Comprehensive numerical
checks confirm that this is indeed the case. 

Using the branching rules (\ref{eq:branching}) we arrive at the generalized
dressing formula for integer spins given in eqn.~(\ref{eq:identity}) where we have fitted the normalization factor numerically.

\section{Derivation of dressing formulas for half-integer spin \label{halfintegerbranching}}
As before we denote the half integer spins as $s+1/2$ where $s$ is integer.
Below we explicitly derive the dressing formulas for $s=1$ and $s=2$ by going through the iterative
procedure described in section~\ref{recursion}.  
We start from the observation made in section~\ref{dressing}  that
\begin{equation}
L((s,s,0,0)|0)\otimes V_{R}(1/2)=V_{R}(s+1/2)\oplus V^{(2)},
\end{equation}
where $V^{(2)}$ has to be expressed in terms of irreps.

\subsection{Formulas for $s=1$}

Let us start with $s=1$. The states of $V^{(2)}$ in the $\mathfrak{o}_{4}$
weight space are
\begin{equation}
\begin{array}{ccc}
 & 1\\
1 & 3 & 1\\
1 & 3 & 1\\
 & 1
\end{array}-\begin{array}{ccc}
 & 1\\
 & 1\\
 & 1\\
 & 1
\end{array}=\begin{array}{ccc}
\\
1 & 2 & 1\\
1 & 2 & 1\\
\\
\end{array}.
\end{equation}
The state $(1,1/2)\equiv(0,1,0,1)\otimes|+\rangle$ is a highest weight state
in the factor space $V^{(2)}$ for which
\begin{equation}
\begin{split}\mu_{4}^{(2)}(u)= & \frac{u-1}{u+1/2},\quad\mu_{3}^{(2)}(u)=\frac{(u-1)(u+1)^{2}}{u^{2}(u+1/2)},\\
P_{1}^{(2)}(u)= & \frac{u^{2}}{(u+1)^{2}},\quad P_{2}^{(2)}(u)=\frac{(u-1)(u+1/2)}{(u+1)(u-1/2)},
\end{split}
\end{equation}
which agree with the highest weights of the representation $L((1,0,0,0)|1)\otimes V_{L}(1/2)$.
Since the dimensions of $L((1,0,0,0)|1)\otimes V_{L}(1/2)$ and $V^{(2)}$
agree
\begin{equation}
L((1,1,0,0)|0)\otimes V_{R}(1/2)=V_{R}(3/2)\oplus\left[L((1,0,0,0)|1)\otimes V_{L}(1/2)\right].
\end{equation}
The corresponding generalized dressing is
\begin{equation}
\langle \mbox{MPS}_{4}^{R}|=\frac{1}{Q_{\theta}(1/2)}\langle\sigma^{R}|\mathcal{T}_{(1,1,0,0)}(0)-\frac{Q_{\theta}(3/2)}{Q_{\theta}(1/2)^{2}}\langle\sigma^{L}|\mathcal{T}_{(1,0,0,0)}(-1).
\end{equation}

\subsection{Formulas for $s=2$}

The states of $V^{(2)}$ in the $\mathfrak{o}_{4}$ weight space are
\begin{equation}
\begin{array}{ccccc}
 &  & 1\\
 & 1 & 3 & 1\\
1 & 3 & 6 & 3 & 1\\
1 & 3 & 6 & 3 & 1\\
 & 1 & 3 & 1\\
 &  & 1
\end{array}-\begin{array}{ccccc}
 &  & 1\\
 &  & 1\\
 &  & 1\\
 &  & 1\\
 &  & 1\\
 &  & 1
\end{array}=\begin{array}{ccccc}
\\
 & 1 & 2 & 1\\
1 & 3 & 5 & 3 & 1\\
1 & 3 & 5 & 3 & 1\\
 & 1 & 2 & 1\\
\\
\end{array}.
\end{equation}
The state $(1,3/2)\equiv(0,1,1,2)\otimes|+\rangle$ is highest weight
in the factor space $V^{(2)}$ for which
\begin{equation}
\begin{split}\mu_{4}^{(2)}(u)= & \frac{u-2}{u+1/2},\quad\mu_{3}^{(2)}(u)=\frac{(u-2)(u+1)^{2}}{u^{2}(u+1/2)},\\
P_{1}^{(2)}(u)= & \frac{u^{2}}{(u+1)^{2}},\quad P_{2}^{(2)}(u)=\frac{(u+2)(u-1)^{2}(u+1/2)}{(u-2)(u+1)^{2}(u-1/2)},
\end{split}
\end{equation}
which agree with the highest weights of the representation $L((2,1,0,0)|2)\otimes V_{L}(1/2)$, therefore
\begin{equation}
L((2,1,0,0)|2)\otimes V_{L}(1/2)=V^{(2)}\oplus V^{(3)}.
\end{equation}
The states of $V^{(3)}$ in the $\mathfrak{o}_{4}$ weight space are
\begin{equation}
\begin{array}{ccccc}
 & 1 & 2 & 1\\
1 & 4 & 6 & 4 & 1\\
1 & 4 & 6 & 4 & 1\\
 & 1 & 2 & 1
\end{array}-\begin{array}{ccccc}
 & 1 & 2 & 1\\
1 & 3 & 5 & 3 & 1\\
1 & 3 & 5 & 3 & 1\\
 & 1 & 2 & 1
\end{array}=\begin{array}{ccccc}
\\
 & 1 & 1 & 1\\
 & 1 & 1 & 1\\
\\
\end{array}.
\end{equation}
The $\mathfrak{o}_{4}$ weight $(1,1/2)$ corresponds to the 4-dimensional
subspace which is spanned by the vectors $|\Lambda_{1}\rangle\otimes|-\rangle$,
$|\Lambda_{2}\rangle\otimes|+\rangle$, $|\Lambda_{3}\rangle\otimes|+\rangle$
and $|\Lambda_{4}\rangle\otimes|+\rangle$ where
\begin{equation}
\Lambda_{1}=\begin{array}{ccccccc}
2 &  & 1 &  & 0 &  & 0\\
 & 1 &  & 0 &  & 0\\
 &  & 1 &  & 0\\
 &  &  & 0
\end{array},\quad\Lambda_{2}=\begin{array}{ccccccc}
2 &  & 1 &  & 0 &  & 0\\
 & 1 &  & 1 &  & 0\\
 &  & 1 &  & 0\\
 &  &  & 0
\end{array},\quad\Lambda_{3}=\begin{array}{ccccccc}
2 &  & 1 &  & 0 &  & 0\\
 & 2 &  & 0 &  & 0\\
 &  & 1 &  & 0\\
 &  &  & 0
\end{array},\quad\Lambda_{4}=\begin{array}{ccccccc}
2 &  & 1 &  & 0 &  & 0\\
 & 1 &  & 0 &  & 0\\
 &  & 1 &  & 0\\
 &  &  & 1
\end{array}.
\end{equation}
The vector 
\begin{equation}
2|\Lambda_{1}\rangle\otimes|-\rangle-|\Lambda_{2}\rangle\otimes|+\rangle+|\Lambda_{3}\rangle\otimes|+\rangle+|\Lambda_{4}\rangle\otimes|+\rangle,
\end{equation}
is a highest weight state for which
\begin{equation}
\begin{split}\mu_{4}^{(3)}(u)= & \frac{u(u+1)}{(u+1/2)(u+2)},\quad\mu_{3}^{(3)}(u)=\frac{u+1}{u+1/2},\\
P_{1}^{(3)}(u)= & \frac{u}{u+2},\quad P_{2}^{(3)}(u)=\frac{(u-1)(u+1/2)}{(u+1)(u-1/2)},
\end{split}
\end{equation}
which agree with the highest weights of the representation $L((1,0,0,0)|2)\otimes V_{L}(1/2)$,
therefore 
\begin{equation}
L((1,0,0,0)|2)\otimes V_{L}(1/2)=V^{(3)}\oplus V_{R}(1/2).
\end{equation}
 The corresponding generalized dressing is
\begin{multline}
\langle \mbox{MPS}_{6}^{R}|=\frac{1}{Q_{\theta}(1/2)}\langle\sigma^{R}|\mathcal{T}_{(2,2,0,0)}(0)+\\
+\frac{Q_{\theta}(5/2)}{Q_{\theta}(1/2)^{2}}\left(\langle\sigma^{L}|\mathcal{T}_{(1,0,0,0)}(-2)-\langle\sigma^{L}|\mathcal{T}_{(2,1,0,0)}(-2)-\langle\sigma^{R}|\mathcal{T}_{(0,0,0,0)}(-2)\right),
\end{multline}
where the normalization factor was determined numerically.

\subsection{Formulas for general $s$}

Based on the previous analysis we have the following conjecture for the representation 
embeddings 
\begin{equation}
\begin{split}L((s,s,0,0)|0)\otimes V_{R}(1/2) & =V_{R}(s+1/2)\oplus V^{(2)},\\
L((s,s-1,0,0)|s)\otimes V_{L}(1/2) & =V^{(2)}\oplus V^{(3)},\\
L((s-1,s-2,0,0)|s)\otimes V_{L}(1/2) & =V^{(3)}\oplus\left[L((s-2,s-2,0,0)|s)\otimes V_{R}(1/2)\right].
\end{split}
\label{eq:branching-1}
\end{equation}
We notice that the dimensions are consistent. We validated the formulas numerically.
Using the branching rules (\ref{eq:branching-1}) we arrive at the generalized
dressing formula for half-integer spins given in eqn.~(\ref{eq:halfintdressing}) where again the normalization factor was
fitted  numerically.

\bibliographystyle{elsarticle-num}
\bibliography{ref}

\begin{thebibliography}{10}
\expandafter\ifx\csname url\endcsname\relax
  \def\url#1{\texttt{#1}}\fi
\expandafter\ifx\csname urlprefix\endcsname\relax\def\urlprefix{URL }\fi
\expandafter\ifx\csname href\endcsname\relax
  \def\href#1#2{#2} \def\path#1{#1}\fi

\bibitem{deLeeuw:2015hxa}
M.~de~Leeuw, C.~Kristjansen, K.~Zarembo, {One-point Functions in Defect CFT and
  Integrability}, JHEP 08 (2015) 098.
\newblock \href {http://arxiv.org/abs/1506.06958} {\path{arXiv:1506.06958}},
  \href {https://doi.org/10.1007/JHEP08(2015)098}
  {\path{doi:10.1007/JHEP08(2015)098}}.

\bibitem{deLeeuw:2016ofj}
M.~de~Leeuw, C.~Kristjansen, G.~Linardopoulos, {One-point functions of
  non-protected operators in the SO(5) symmetric D3\textendash{}D7 dCFT}, J.
  Phys. A 50~(25) (2017) 254001.
\newblock \href {http://arxiv.org/abs/1612.06236} {\path{arXiv:1612.06236}},
  \href {https://doi.org/10.1088/1751-8121/aa714b}
  {\path{doi:10.1088/1751-8121/aa714b}}.

\bibitem{Kristjansen:2021abc}
C.~Kristjansen, D.-L. Vu, K.~Zarembo, {Integrable domain walls in ABJM theory},
  JHEP 02 (2022) 070.
\newblock \href {http://arxiv.org/abs/2112.10438} {\path{arXiv:2112.10438}},
  \href {https://doi.org/10.1007/JHEP02(2022)070}
  {\path{doi:10.1007/JHEP02(2022)070}}.

\bibitem{Kristjansen:2023ysz}
C.~Kristjansen, K.~Zarembo, {\textquoteright{}t Hooft loops and integrability},
  JHEP 08 (2023) 184.
\newblock \href {http://arxiv.org/abs/2305.03649} {\path{arXiv:2305.03649}},
  \href {https://doi.org/10.1007/JHEP08(2023)184}
  {\path{doi:10.1007/JHEP08(2023)184}}.

\bibitem{deLeeuw:2024qki}
M.~de~Leeuw, A.~Holguin, {Integrable Conformal Defects in N=4 SYM} (6 2024).
\newblock \href {http://arxiv.org/abs/2406.13741} {\path{arXiv:2406.13741}}.

\bibitem{Minahan:2002ve}
J.~A. Minahan, K.~Zarembo, {The Bethe ansatz for N=4 superYang-Mills}, JHEP 03
  (2003) 013.
\newblock \href {http://arxiv.org/abs/hep-th/0212208}
  {\path{arXiv:hep-th/0212208}}, \href
  {https://doi.org/10.1088/1126-6708/2003/03/013}
  {\path{doi:10.1088/1126-6708/2003/03/013}}.

\bibitem{Minahan:2008hf}
J.~A. Minahan, K.~Zarembo, {The Bethe ansatz for superconformal Chern-Simons},
  JHEP 09 (2008) 040.
\newblock \href {http://arxiv.org/abs/0806.3951} {\path{arXiv:0806.3951}},
  \href {https://doi.org/10.1088/1126-6708/2008/09/040}
  {\path{doi:10.1088/1126-6708/2008/09/040}}.

\bibitem{Ivanovskiy:2024vel}
V.~Ivanovskiy, S.~Komatsu, V.~Mishnyakov, N.~Terziev, N.~Zaigraev, K.~Zarembo,
  {Vacuum Condensates on the Coulomb Branch} (5 2024).
\newblock \href {http://arxiv.org/abs/2405.19043} {\path{arXiv:2405.19043}}.

\bibitem{Jiang:2019xdz}
Y.~Jiang, S.~Komatsu, E.~Vescovi, {Structure constants in $ \mathcal{N} $ = 4
  SYM at finite coupling as worldsheet g-function}, JHEP 07~(07) (2020) 037.
\newblock \href {http://arxiv.org/abs/1906.07733} {\path{arXiv:1906.07733}},
  \href {https://doi.org/10.1007/JHEP07(2020)037}
  {\path{doi:10.1007/JHEP07(2020)037}}.

\bibitem{Yang:2021hrl}
P.~Yang, Y.~Jiang, S.~Komatsu, J.-B. Wu, {Three-point functions in ABJM and
  Bethe Ansatz}, JHEP 01 (2022) 002.
\newblock \href {http://arxiv.org/abs/2103.15840} {\path{arXiv:2103.15840}},
  \href {https://doi.org/10.1007/JHEP01(2022)002}
  {\path{doi:10.1007/JHEP01(2022)002}}.

\bibitem{Yang:2022dlk}
P.~Yang, {Integrable boundary states from maximal giant gravitons in ABJM
  theory}, Phys. Lett. B 846 (2023) 138194.
\newblock \href {http://arxiv.org/abs/2208.12010} {\path{arXiv:2208.12010}},
  \href {https://doi.org/10.1016/j.physletb.2023.138194}
  {\path{doi:10.1016/j.physletb.2023.138194}}.

\bibitem{Kristjansen:2024dnm}
C.~Kristjansen, K.~Zarembo, {Integrable Holographic Defect CFTs}, in: Gravity,
  Fields and Strings, A Conference in Honour of Gordon Semenoff, 2024.
\newblock \href {http://arxiv.org/abs/2401.17144} {\path{arXiv:2401.17144}}.

\bibitem{Zamolodchikov:1989fp}
A.~B. Zamolodchikov, {Integrals of Motion and S Matrix of the (Scaled) T=T(c)
  Ising Model with Magnetic Field}, Int. J. Mod. Phys. A 4 (1989) 4235.
\newblock \href {https://doi.org/10.1142/S0217751X8900176X}
  {\path{doi:10.1142/S0217751X8900176X}}.

\bibitem{Piroli:2017sei}
L.~Piroli, B.~Pozsgay, E.~Vernier, {What is an integrable quench?}, Nucl. Phys.
  B 925 (2017) 362--402.
\newblock \href {http://arxiv.org/abs/1709.04796} {\path{arXiv:1709.04796}},
  \href {https://doi.org/10.1016/j.nuclphysb.2017.10.012}
  {\path{doi:10.1016/j.nuclphysb.2017.10.012}}.

\bibitem{Kristjansen:2020mhn}
C.~Kristjansen, D.~M\"uller, K.~Zarembo, {Integrable boundary states in D3-D5
  dCFT: beyond scalars}, JHEP 08 (2020) 103.
\newblock \href {http://arxiv.org/abs/2005.01392} {\path{arXiv:2005.01392}},
  \href {https://doi.org/10.1007/JHEP08(2020)103}
  {\path{doi:10.1007/JHEP08(2020)103}}.

\bibitem{Pozsgay_2014}
B.~Pozsgay, {Overlaps between eigenstates of the XXZ spin-1/2 chain and a class
  of simple product states}, Journal of Statistical Mechanics: Theory and
  Experiment 2014~(6) (2014) P06011.
\newblock \href {https://doi.org/10.1088/1742-5468/2014/06/p06011}
  {\path{doi:10.1088/1742-5468/2014/06/p06011}}.

\bibitem{Brockmann_2014odd}
M.~Brockmann, J.~De~Nardis, B.~Wouters, J.-S. Caux, Neel-xxz state overlaps:
  odd particle numbers and lieb liniger scaling limit, Journal of Physics A:
  Mathematical and Theoretical 47~(34) (2014) 345003.
\newblock \href {https://doi.org/10.1088/1751-8113/47/34/345003}
  {\path{doi:10.1088/1751-8113/47/34/345003}}.

\bibitem{Brockmann_2014}
M.~Brockmann, J.~De~Nardis, B.~Wouters, J.-S. Caux, A gaudin-like determinant
  for overlaps of neel and xxz bethe states, Journal of Physics A: Mathematical
  and Theoretical 47~(14) (2014) 145003.
\newblock \href {https://doi.org/10.1088/1751-8113/47/14/145003}
  {\path{doi:10.1088/1751-8113/47/14/145003}}.

\bibitem{Gombor:2021uxz}
T.~Gombor, B.~Pozsgay, {On factorized overlaps: Algebraic Bethe Ansatz, twists,
  and Separation of Variables}, Nucl. Phys. B 967 (2021) 115390.
\newblock \href {http://arxiv.org/abs/2101.10354} {\path{arXiv:2101.10354}},
  \href {https://doi.org/10.1016/j.nuclphysb.2021.115390}
  {\path{doi:10.1016/j.nuclphysb.2021.115390}}.

\bibitem{Gombor:2021hmj}
T.~Gombor, {On exact overlaps for gl(N) symmetric spin chains}, Nucl. Phys. B
  983 (2022) 115909.
\newblock \href {http://arxiv.org/abs/2110.07960} {\path{arXiv:2110.07960}},
  \href {https://doi.org/10.1016/j.nuclphysb.2022.115909}
  {\path{doi:10.1016/j.nuclphysb.2022.115909}}.

\bibitem{Gombor:2023bez}
T.~Gombor, {Exact overlaps for all integrable two-site boundary states of $
  \mathfrak{gl} $(N) symmetric spin chains}, JHEP 05 (2024) 194.
\newblock \href {http://arxiv.org/abs/2311.04870} {\path{arXiv:2311.04870}},
  \href {https://doi.org/10.1007/JHEP05(2024)194}
  {\path{doi:10.1007/JHEP05(2024)194}}.

\bibitem{Kristjansen:2020vbe}
C.~Kristjansen, D.~M\"uller, K.~Zarembo, {Overlaps and fermionic dualities for
  integrable super spin chains}, JHEP 03 (2021) 100.
\newblock \href {http://arxiv.org/abs/2011.12192} {\path{arXiv:2011.12192}},
  \href {https://doi.org/10.1007/JHEP03(2021)100}
  {\path{doi:10.1007/JHEP03(2021)100}}.

\bibitem{Buhl-Mortensen:2015gfd}
I.~Buhl-Mortensen, M.~de~Leeuw, C.~Kristjansen, K.~Zarembo, {One-point
  Functions in AdS/dCFT from Matrix Product States}, JHEP 02 (2016) 052.
\newblock \href {http://arxiv.org/abs/1512.02532} {\path{arXiv:1512.02532}},
  \href {https://doi.org/10.1007/JHEP02(2016)052}
  {\path{doi:10.1007/JHEP02(2016)052}}.

\bibitem{DeLeeuw:2018cal}
M.~De~Leeuw, C.~Kristjansen, G.~Linardopoulos, {Scalar one-point functions and
  matrix product states of AdS/dCFT}, Phys. Lett. B 781 (2018) 238--243.
\newblock \href {http://arxiv.org/abs/1802.01598} {\path{arXiv:1802.01598}},
  \href {https://doi.org/10.1016/j.physletb.2018.03.083}
  {\path{doi:10.1016/j.physletb.2018.03.083}}.

\bibitem{deLeeuw:2017cop}
M.~de~Leeuw, A.~C. Ipsen, C.~Kristjansen, M.~Wilhelm, {Introduction to
  integrability and one-point functions in $\mathcal N=$ 4 supersymmetric
  Yang\textendash{}Mills theory and its defect cousin}, Les Houches Lecture
  notes 106 (2019).
\newblock \href {http://arxiv.org/abs/1708.02525} {\path{arXiv:1708.02525}},
  \href {https://doi.org/10.1093/oso/9780198828150.003.0008}
  {\path{doi:10.1093/oso/9780198828150.003.0008}}.

\bibitem{deLeeuw:2019usb}
M.~de~Leeuw, {One-point functions in AdS/dCFT}, J. Phys. A 53~(28) (2020)
  283001.
\newblock \href {http://arxiv.org/abs/1908.03444} {\path{arXiv:1908.03444}},
  \href {https://doi.org/10.1088/1751-8121/ab15fb}
  {\path{doi:10.1088/1751-8121/ab15fb}}.

\bibitem{Linardopoulos:2020jck}
G.~Linardopoulos, {Solving holographic defects}, PoS CORFU2019 (2020) 141.
\newblock \href {http://arxiv.org/abs/2005.02117} {\path{arXiv:2005.02117}},
  \href {https://doi.org/10.22323/1.376.0141} {\path{doi:10.22323/1.376.0141}}.

\bibitem{Pozsgay:2018dzs}
B.~Pozsgay, L.~Piroli, E.~Vernier, {Integrable Matrix Product States from
  boundary integrability}, SciPost Phys. 6~(5) (2019) 062.
\newblock \href {http://arxiv.org/abs/1812.11094} {\path{arXiv:1812.11094}},
  \href {https://doi.org/10.21468/SciPostPhys.6.5.062}
  {\path{doi:10.21468/SciPostPhys.6.5.062}}.

\bibitem{DeLeeuw:2019ohp}
M.~De~Leeuw, T.~Gombor, C.~Kristjansen, G.~Linardopoulos, B.~Pozsgay, {Spin
  Chain Overlaps and the Twisted Yangian}, JHEP 01 (2020) 176.
\newblock \href {http://arxiv.org/abs/1912.09338} {\path{arXiv:1912.09338}},
  \href {https://doi.org/10.1007/JHEP01(2020)176}
  {\path{doi:10.1007/JHEP01(2020)176}}.

\bibitem{Molev:1997wp}
A.~Molev, {Finite dimensional irreducible representations of twisted Yangians},
  J. Math. Phys. 39 (1998) 5559--5600.
\newblock \href {http://arxiv.org/abs/q-alg/9711022}
  {\path{arXiv:q-alg/9711022}}, \href {https://doi.org/10.1063/1.532551}
  {\path{doi:10.1063/1.532551}}.

\bibitem{Doikou:2004hy}
A.~Doikou, {On reflection algebras and boundary Yangians}, J. Math. Phys. 46
  (2005) 053504.
\newblock \href {http://arxiv.org/abs/hep-th/0403277}
  {\path{arXiv:hep-th/0403277}}, \href {https://doi.org/10.1063/1.1888029}
  {\path{doi:10.1063/1.1888029}}.

\bibitem{MacKay:2004tc}
N.~J. MacKay, {Introduction to Yangian symmetry in integrable field theory},
  Int. J. Mod. Phys. A 20 (2005) 7189--7218.
\newblock \href {http://arxiv.org/abs/hep-th/0409183}
  {\path{arXiv:hep-th/0409183}}, \href
  {https://doi.org/10.1142/S0217751X05022317}
  {\path{doi:10.1142/S0217751X05022317}}.

\bibitem{Zakharov:1973pp}
V.~E. Zakharov, A.~V. Mikhailov, {Relativistically Invariant Two-Dimensional
  Models in Field Theory Integrable by the Inverse Problem Technique. (In
  Russian)}, Sov. Phys. JETP 47 (1978) 1017--1027.

\bibitem{Zakharov:1980ty}
V.~E. Zakharov, A.~V. Mikhailov, {ON THE INTEGRABILITY OF CLASSICAL SPINOR
  MODELS IN TWO-DIMENSIONAL SPACE-TIME}, Commun. Math. Phys. 74 (1980) 21--40.
\newblock \href {https://doi.org/10.1007/BF01197576}
  {\path{doi:10.1007/BF01197576}}.

\bibitem{Harnad:1983we}
J.~P. Harnad, Y.~Saint~Aubin, S.~Shnider, {Backlund Transformations for
  Nonlinear $\sigma$ Models With Values in Riemannian Symmetric Spaces},
  Commun. Math. Phys. 92 (1984) 329.
\newblock \href {https://doi.org/10.1007/BF01210726}
  {\path{doi:10.1007/BF01210726}}.

\bibitem{Spradlin:2006wk}
M.~Spradlin, A.~Volovich, {Dressing the Giant Magnon}, JHEP 10 (2006) 012.
\newblock \href {http://arxiv.org/abs/hep-th/0607009}
  {\path{arXiv:hep-th/0607009}}, \href
  {https://doi.org/10.1088/1126-6708/2006/10/012}
  {\path{doi:10.1088/1126-6708/2006/10/012}}.

\bibitem{Kalousios:2006xy}
C.~Kalousios, M.~Spradlin, A.~Volovich, {Dressing the giant magnon II}, JHEP 03
  (2007) 020.
\newblock \href {http://arxiv.org/abs/hep-th/0611033}
  {\path{arXiv:hep-th/0611033}}, \href
  {https://doi.org/10.1088/1126-6708/2007/03/020}
  {\path{doi:10.1088/1126-6708/2007/03/020}}.

\bibitem{Jevicki:2007pk}
A.~Jevicki, C.~Kalousios, M.~Spradlin, A.~Volovich, {Dressing the Giant Gluon},
  JHEP 12 (2007) 047.
\newblock \href {http://arxiv.org/abs/0708.0818} {\path{arXiv:0708.0818}},
  \href {https://doi.org/10.1088/1126-6708/2007/12/047}
  {\path{doi:10.1088/1126-6708/2007/12/047}}.

\bibitem{Constable:2001ag}
N.~R. Constable, R.~C. Myers, O.~Tafjord, {NonAbelian brane intersections},
  JHEP 06 (2001) 023.
\newblock \href {http://arxiv.org/abs/hep-th/0102080}
  {\path{arXiv:hep-th/0102080}}, \href
  {https://doi.org/10.1088/1126-6708/2001/06/023}
  {\path{doi:10.1088/1126-6708/2001/06/023}}.

\bibitem{Myers:2008me}
R.~C. Myers, M.~C. Wapler, {Transport Properties of Holographic Defects}, JHEP
  12 (2008) 115.
\newblock \href {http://arxiv.org/abs/0811.0480} {\path{arXiv:0811.0480}},
  \href {https://doi.org/10.1088/1126-6708/2008/12/115}
  {\path{doi:10.1088/1126-6708/2008/12/115}}.

\bibitem{deLeeuw:2016umh}
M.~de~Leeuw, C.~Kristjansen, S.~Mori, {AdS/dCFT one-point functions of the
  SU(3) sector}, Phys. Lett. B 763 (2016) 197--202.
\newblock \href {http://arxiv.org/abs/1607.03123} {\path{arXiv:1607.03123}},
  \href {https://doi.org/10.1016/j.physletb.2016.10.044}
  {\path{doi:10.1016/j.physletb.2016.10.044}}.

\bibitem{Gombor:2020kgu}
T.~Gombor, Z.~Bajnok, {Boundary states, overlaps, nesting and bootstrapping
  AdS/dCFT}, JHEP 10 (2020) 123.
\newblock \href {http://arxiv.org/abs/2004.11329} {\path{arXiv:2004.11329}},
  \href {https://doi.org/10.1007/JHEP10(2020)123}
  {\path{doi:10.1007/JHEP10(2020)123}}.

\bibitem{Gombor:2020auk}
T.~Gombor, Z.~Bajnok, {Boundary state bootstrap and asymptotic overlaps in
  AdS/dCFT}, JHEP 03 (2021) 222.
\newblock \href {http://arxiv.org/abs/2006.16151} {\path{arXiv:2006.16151}},
  \href {https://doi.org/10.1007/JHEP03(2021)222}
  {\path{doi:10.1007/JHEP03(2021)222}}.

\bibitem{Gombor:2022aqj}
T.~Gombor, C.~Kristjansen, {Overlaps for matrix product states of arbitrary
  bond dimension in ABJM theory}, Phys. Lett. B 834 (2022) 137428.
\newblock \href {http://arxiv.org/abs/2207.06866} {\path{arXiv:2207.06866}},
  \href {https://doi.org/10.1016/j.physletb.2022.137428}
  {\path{doi:10.1016/j.physletb.2022.137428}}.

\bibitem{molev2002gelfandtsetlinbasesclassicallie}
A.~I. Molev, Gelfand-tsetlin bases for classical lie algebras (2002).
\newblock \href {http://arxiv.org/abs/math/0211289}
  {\path{arXiv:math/0211289}}.

\bibitem{Gombor:2019bun}
T.~Gombor, {On the classification of rational K-matrices}, J. Phys. A 53~(13)
  (2020) 135203.
\newblock \href {http://arxiv.org/abs/1904.03044} {\path{arXiv:1904.03044}},
  \href {https://doi.org/10.1088/1751-8121/ab7602}
  {\path{doi:10.1088/1751-8121/ab7602}}.

\bibitem{Tsuboi:1998ne}
Z.~Tsuboi, {Analytic Bethe Ansatz And Functional Equations Associated With Any
  Simple Root Systems Of The Lie Superalgebra $sl(r+1|s+1)$}, Physica A 252
  (1998) 565--585.
\newblock \href {http://arxiv.org/abs/0911.5387} {\path{arXiv:0911.5387}},
  \href {https://doi.org/10.1016/S0378-4371(97)00625-0}
  {\path{doi:10.1016/S0378-4371(97)00625-0}}.

\bibitem{Gromov:2019icz}
N.~Gromov, {Spectrum of $\mathcal N=$ 4 supersymmetric Yang\textendash{}Mills
  theory and the quantum spectral curve} (2019).
\newblock \href {https://doi.org/10.1093/oso/9780198828150.003.0009}
  {\path{doi:10.1093/oso/9780198828150.003.0009}}.

\bibitem{Kazakov:2018ugh}
V.~Kazakov, {Quantum Spectral Curve of $\gamma$-twisted ${\cal N}=4$ SYM theory
  and fishnet CFT} (2018) 293--342\href {http://arxiv.org/abs/1802.02160}
  {\path{arXiv:1802.02160}}, \href {https://doi.org/10.1142/9789813233867_0016}
  {\path{doi:10.1142/9789813233867_0016}}.

\bibitem{Levkovich-Maslyuk:2019awk}
F.~Levkovich-Maslyuk, {A review of the AdS/CFT Quantum Spectral Curve}, J.
  Phys. A 53~(28) (2020) 283004.
\newblock \href {http://arxiv.org/abs/1911.13065} {\path{arXiv:1911.13065}},
  \href {https://doi.org/10.1088/1751-8121/ab7137}
  {\path{doi:10.1088/1751-8121/ab7137}}.

\bibitem{Sklyanin:1995bm}
E.~K. Sklyanin, {Separation of variables - new trends}, Prog. Theor. Phys.
  Suppl. 118 (1995) 35--60.
\newblock \href {http://arxiv.org/abs/solv-int/9504001}
  {\path{arXiv:solv-int/9504001}}, \href {https://doi.org/10.1143/PTPS.118.35}
  {\path{doi:10.1143/PTPS.118.35}}.

\bibitem{Gromov:2016itr}
N.~Gromov, F.~Levkovich-Maslyuk, G.~Sizov, {New Construction of Eigenstates and
  Separation of Variables for SU(N) Quantum Spin Chains}, JHEP 09 (2017) 111.
\newblock \href {http://arxiv.org/abs/1610.08032} {\path{arXiv:1610.08032}},
  \href {https://doi.org/10.1007/JHEP09(2017)111}
  {\path{doi:10.1007/JHEP09(2017)111}}.

\bibitem{Ekhammar:2023iph}
S.~Ekhammar, N.~Gromov, P.~Ryan, {Boundary overlaps from Functional Separation
  of Variables}, JHEP 05 (2024) 268.
\newblock \href {http://arxiv.org/abs/2312.11612} {\path{arXiv:2312.11612}},
  \href {https://doi.org/10.1007/JHEP05(2024)268}
  {\path{doi:10.1007/JHEP05(2024)268}}.

\bibitem{Kristjansen:2012tn}
C.~Kristjansen, G.~W. Semenoff, D.~Young, {Chiral primary one-point functions
  in the D3-D7 defect conformal field theory}, JHEP 01 (2013) 117.
\newblock \href {http://arxiv.org/abs/1210.7015} {\path{arXiv:1210.7015}},
  \href {https://doi.org/10.1007/JHEP01(2013)117}
  {\path{doi:10.1007/JHEP01(2013)117}}.

\bibitem{deLeeuw:2019sew}
M.~de~Leeuw, C.~Kristjansen, K.~E. Vardinghus, {A non-integrable quench from
  AdS/dCFT}, Phys. Lett. B 798 (2019) 134940.
\newblock \href {http://arxiv.org/abs/1906.10714} {\path{arXiv:1906.10714}},
  \href {https://doi.org/10.1016/j.physletb.2019.134940}
  {\path{doi:10.1016/j.physletb.2019.134940}}.

\bibitem{Gombor:2024iix}
T.~Gombor, {Exact overlaps for ''all'' integrable matrix product states of
  rational spin chains} (10 2024).
\newblock \href {http://arxiv.org/abs/2410.23282} {\path{arXiv:2410.23282}}.

\end{thebibliography}

\end{document}